\documentclass[11pt]{article}

\usepackage{graphicx,amsmath,amsfonts,amssymb,dcolumn,amsthm,euscript}

\begin{document}

\title{\textbf{A local and renormalizable framework for the  gauge-invariant  operator 
$A^2_{\min}$ in Euclidean Yang-Mills theories in linear covariant gauges
}}
\author{\textbf{M.~A.~L.~Capri}\thanks{caprimarcio@gmail.com}, 
\textbf{D.~Fiorentini} \thanks{diegofiorentinia@gmail.com}, 
\textbf{M.~S.~Guimaraes}
\thanks{msguimaraes@uerj.br}, \\ 
\textbf{B.~W.~Mintz}\thanks{bruno.mintz@uerj.br}, 
\textbf{L.~F.~Palhares} \thanks{leticia.palhares@uerj.br}, 
\textbf{S.~P.~Sorella}\thanks{silvio.sorella@gmail.com}, \\\\\
\textit{{\small UERJ -- Universidade do Estado do Rio de Janeiro,}}\\
\textit{{\small Instituto de F\'isica -- Departamento de F\'{\i}sica Te\'orica -- Rua S\~ao Francisco Xavier 524,}}\\
\textit{{\small 20550-013, Maracan\~a, Rio de Janeiro, Brasil}}}
\date{}
\maketitle

\begin{abstract}
We address the issue of the renormalizability of the gauge-invariant non-local dimension-two operator 
$A^2_{\rm min}$,  whose minimization is defined along the gauge orbit.  Despite its non-local 
character, we show that the operator $A^2_{\rm min}$ 
can be cast in local form through the introduction of an auxiliary Stueckelberg field. The 
localization procedure gives rise to an unconventional kind of Stueckelberg-type action which 
turns out to be renormalizable to all orders of perturbation theory. In particular, as a 
consequence of its gauge invariance, the anomalous dimension of the operator $A^2_{\rm min}$ 
turns out to be independent from the gauge parameter $\alpha$ entering the gauge-fixing 
condition, being thus given by the anomalous dimension of the operator $A^2$ in the Landau 
gauge.

\end{abstract}

\section{Introduction}

%

Dimension-two condensates have been object of intensive investigations in recent years. These 
condensates might play an important
role in the non-perturbative regime of Euclidean Yang-Mills theories, as pointed out  by the 
considerable amount of results obtained through
theoretical and phenomenological studies as well as from lattice
simulations \cite
{Cornwall:1981zr,Greensite:1985vq,Stingl:1985hx,Lavelle:1988eg,Gubarev:2000nz,Gubarev:2000eu,
Verschelde:2001ia,Kondo:2001nq, Kondo:2001tm,Dudal:2003vv,
Browne:2003uv,Dudal:2003gu, Dudal:2003by,
Dudal:2004rx,Browne:2004mk, Gracey:2004bk,
Li:2004te,Boucaud:2001st,Boucaud:2002nc,
Boucaud:2005rm, RuizArriola:2004en,Suzuki:2004dw,Gubarev:2005it,Furui:2005bu,Boucaud:2005xn,
Chernodub:2005gz,Boucaud:2005xn,Boucaud:2008gn,Pene:2011kg,Boucaud:2010gr,Blossier:2010ky, Dudal:2010tf,
Boucaud:2011eh,Blossier:2011tf,Blossier:2013te}.\\\\
For instance, the gluon condensate $\left\langle A_{\mu }^{a}A_{\mu
}^{a}\right\rangle $ has been largely investigated in the Landau gauge. As
pointed out in \cite{Gubarev:2000nz}, this condensate enters the operator
product expansion (OPE) of the gluon propagator. Moreover, a combined OPE
and lattice analysis has shown that this condensate can account for the $%
1/Q^{2}$ corrections which have been reported \cite
{Boucaud:2001st,Boucaud:2002nc,Boucaud:2005rm,RuizArriola:2004en,Furui:2005bu,
Boucaud:2005xn,Boucaud:2008gn,Pene:2011kg,Boucaud:2010gr,Blossier:2010ky,Boucaud:2011eh,
Blossier:2011tf,Blossier:2013te}
in the running of the coupling constant and in the gluon
correlation functions. \\\\
An effective potential for $\left\langle
A_{\mu }^{a}A_{\mu }^{a}\right\rangle $ in Landau gauge has been obtained and
evaluated in analytic form at two loops in \cite
{Verschelde:2001ia,Dudal:2003vv,Browne:2003uv,Browne:2004mk,Gracey:2004bk},
showing that a nonvanishing value of $\left\langle A_{\mu
}^{a}A_{\mu }^{a}\right\rangle $ is favoured as it lowers the
vacuum energy. As a consequence, a dynamical gluon mass is
generated. We also recall that, in the Landau gauge, the operator
$A_{\mu }^{a}A_{\mu }^{a}$ is $BRST$-invariant on shell, a
property which has allowed for an all-orders proof of its
multiplicative renormalizability  \cite{Dudal:2002pq}. Its anomalous dimension is not
an independent parameter, being expressed as a combination of the
gauge $\beta - $function and of the anomalous dimension of the gauge field $A_{\mu }^{a}$ 
\cite{Dudal:2002pq}, namely
\begin{equation}
\gamma_{A^{2}}\Big|_{\rm Landau} = \left(  \frac{\beta(a)}{a} + \gamma^{\rm Landau}_{A}(a)   \right)  \;, \qquad a = \frac{g^2}{16\pi^2}    \;,  \label{adl}
\end{equation}
where $(\beta(a), \gamma^{\rm Landau}_A(a))$ denote, respectively, the $\beta$-function and 
the anomalous dimension of the gauge field $A_\mu$ in the Landau gauge.
This
relation was conjectured and explicitly verified up to three-loop
order in \cite{Gracey:2002yt}. \\\\
Dimension-two condensates also play an important role within 
the context of the Gribov-Zwanziger approach to confinement 
\cite{Dudal:2005na,Dudal:2007cw,Dudal:2008sp,Dudal:2011gd,Vandersickel:2012tz}  as well as for the 
formation of a dynamical gluon mass within the framework of the Dyson-Schwinger equations in Landau 
gauge, as reported in \cite
{Cornwall:1981zr, Aguilar:2008xm,Aguilar:2015bud}. These non-perturbative effects give rise to the 
so called decoupling solution for the gluon propagator 
\cite{Cornwall:1981zr,Dudal:2005na,Dudal:2007cw,Dudal:2008sp,Aguilar:2008xm,Fischer:2008uz}, 
{\it i.e.} to a propagator which exhibits positivity violation, while attaining a finite non-vanishing 
value at zero momentum. Until now, this behaviour is in very good agreement with the most recent lattice 
numerical simulations  \cite{Cucchieri:2007md,Cucchieri:2007rg,Cucchieri:2011ig,Oliveira:2012eh}. The  
generalization of these results to the linear covariant gauges has been worked out recently and can be found in  
\cite{Sobreiro:2005vn,Aguilar:2015nqa,Huber:2015ria,Capri:2015pja,Capri:2015ixa,Capri:2015nzw,Cucchieri:2009kk,Cucchieri:2011aa,Bicudo:2015rma}. \\\\Despite the huge amount of results obtained so far, it seems fair to state that many aspects related to dimension-two operators deserve a better understanding. This is certainly the case 
of the gauge invariance, a central issue in order to give a precise physical meaning to the corresponding 
condensates. This is precisely the topic which will be studied in the present work. Let us proceed thus by briefly introducing the genuine gauge-invariant dimension-two operator  $A_{\min }^{2}$.

\subsection{Construction and properties of the operator $A_{\min }^{2}$}
 The gauge-invariant dimension-two operator $A_{\min }^{2}$\footnote {See Appendix \eqref{apb} for more details.} is constructed by minimizing  the functional
$\mathrm{Tr}\int d^{4}x\,A_{\mu }^{u}A_{\mu }^{u}$ along the gauge
orbit of $A_{\mu }$ \cite
{Zwanziger:1990tn,Dell'Antonio:1989jn,Dell'Antonio:1991xt,vanBaal:1991zw}%
, namely
\begin{eqnarray}
A_{\min }^{2} &\equiv &\min_{\{u\}}\mathrm{Tr}\int d^{4}x\,A_{\mu
}^{u}A_{\mu }^{u}\;,
\nonumber \\
A_{\mu }^{u} &=&u^{\dagger }A_{\mu }u+\frac{i}{g}u^{\dagger }\partial _{\mu
}u\;.  \label{Aminn0}
\end{eqnarray}
In particular, the stationary condition of the functional \eqref{Aminn0}  gives rise to a non-local transverse 
field configuration $A^h_\mu$, $\partial_\mu A^h_\mu=0$, which can be expressed as an infinite series in 
the gauge field $A_\mu$, {\it i.e.}
\begin{eqnarray}
A_{\mu }^{h} &=&\left( \delta _{\mu \nu }-\frac{\partial _{\mu }\partial
_{\nu }}{\partial ^{2}}\right) \phi _{\nu }\;,  \qquad  \partial_\mu A^h_\mu= 0 \;, \nonumber \\
\phi _{\nu } &=&A_{\nu }-ig\left[ \frac{1}{\partial ^{2}}\partial A,A_{\nu
}\right] +\frac{ig}{2}\left[ \frac{1}{\partial ^{2}}\partial A,\partial
_{\nu }\frac{1}{\partial ^{2}}\partial A\right] +O(A^{3})\;.  \label{min0}
\end{eqnarray}
Remarkably, the configuration $A_{\mu }^{h}$ turns out to be left invariant
by infinitesimal gauge transformations order by order in the gauge
coupling $g$ \cite{Lavelle:1995ty} (see also Appendix \ref{apb}) as
\begin{eqnarray}
\delta A_{\mu }^{h} &=&0\;,  \nonumber \\
\delta A_{\mu } &=&-\partial _{\mu }\omega +ig\left[ A_{\mu },\omega \right]
\;.  \label{gio}
\end{eqnarray}
Thus, from expression (\ref{Aminn0}) it follows that
\begin{eqnarray}
A_{\min }^{2} &=&\mathrm{Tr}\int d^{4}x\,A_{\mu }^{h}A_{\mu }^{h}\;,  \nonumber \\
&=&\frac{1}{2}\int d^{4}x\left[ A_{\mu }^{a}\left( \delta _{\mu \nu }-\frac{%
\partial _{\mu }\partial _{\nu }}{\partial ^{2}}\right) A_{\nu
}^{a}-gf^{abc}\left( \frac{\partial _{\nu }}{\partial ^{2}}\partial
A^{a}\right) \left( \frac{1}{\partial ^{2}}\partial {A}^{b}\right) A_{\nu
}^{c}\right] \;+O(A^{4})\;.  \label{min1}
\end{eqnarray}
The gauge-invariant nature of expression \eqref{min1} can be made manifest by rewriting it in terms of the  field
strength $F_{\mu \nu }$. In fact, as proven in
\cite{Zwanziger:1990tn}, it turns out that
\begin{eqnarray}
A_{\min }^{2} &=&-\frac{1}{2}\mathrm{Tr}\int d^{4}x\left( F_{\mu
\nu }\frac{1}{D^{2}}F_{\mu \nu }+2i\frac{1}{D^{2}}F_{\lambda \mu
}\left[ \frac{1}{D^{2}}D_{\kappa }F_{\kappa \lambda
},\frac{1}{D^{2}}D_{\nu }F_{\nu \mu }\right] \right.
\nonumber \\
&&-2i\left. \frac{1}{D^{2}}F_{\lambda \mu }\left[ \frac{1}{D^{2}}D_{\kappa
}F_{\kappa \nu },\frac{1}{D^{2}}D_{\nu }F_{\lambda \mu }\right] \right)
+O(F^{4})\;,  \label{zzw}
\end{eqnarray}
from which the gauge invariance becomes apparent. The operator $({D^{2}})^{-1}$ in expression 
\eqref{zzw} denotes the inverse of the Laplacian $D^2=D_\mu D_\mu$ with $D_\mu$ being the 
covariant derivative \cite{Zwanziger:1990tn}. Let us also underline that, in the Landau gauge 
$\partial_\mu A_\mu=0$, the operator $(A^h_\mu A^h_\mu)$ reduces to the operator $A^{2}$
\begin{equation}
(A^{h,a}_\mu A^{h,a}_\mu) \Big|_{\rm Landau} =  A^a_\mu A^a_\mu \;.      \label{land}
\end{equation}

\subsection{Aim of the paper and its structure}

 As already mentioned, the main aim of the present work is to face the issue of the gauge invariance 
 of non-Abelian gauge theories in the presence of dimension-two operators. 
More precisely, we shall provide a 
general and detailed analysis of the gauge-invariant quantity $(A^h_\mu A^h_\mu)$, eq.\eqref{min1}, 
within the framework of 
Euclidean Yang-Mills theories quantized in the class of the linear covariant gauges. We  shall be able to show that, despite 
its non-local character, the operator $(A^h_\mu A^h_\mu)$ can be localized by means of the introduction of an 
auxiliary Stueckelberg field. Nevertheless, the resulting theory can be seen as a kind of unconventional 
Stueckelberg model which does not suffer from the known drawbacks, {\it i.e.}  the non-renormalizability, of 
the usual Stueckelberg mass term. 
 Therefore, we end up with a well-defined framework accounting for the existence of a gauge-invariant dimension-two operator. 
\\\\Relying on an exact BRST invariance, we shall  establish the multiplicative renormalizability of the 
operator $(A^h_\mu A^h_\mu)$ to all orders of perturbation 
theory by means of the algebraic renormalization. 
Moreover, the anomalous
dimension of $ (A^h_\mu A^h_\mu)$ can be proven to be independent from the gauge parameter $\alpha$ and turns out to 
be equal to the anomalous dimension of the operator $A^2$ in the Landau gauge, namely
\begin{equation}
\gamma_{(A^h)^{2}} = \gamma_{A^{2}}\Big|_{\rm Landau} = 
\left(  \frac{\beta(a)}{a} + \gamma^{\rm Landau}_{A}(a)   \right)  \;, \qquad a = \frac{g^2}{16\pi^2}    \;.  \label{ad1}
\end{equation}
We underline that expression \eqref{ad1} is valid to all orders of perturbation theory, thereby extending the 
previous one-loop results obtained in \cite{Gracey:2007ki}. \\\\The paper is organized as follows. In Section 2 
we present the localization procedure for the operator $(A^h_\mu A^h_\mu) $ within the framework of a BRST-invariant 
action. In Section 3 we derive the Ward identities and we establish the all-order renormalizability of 
$(A^h_\mu A^h_\mu)$ by means of the algebraic renormalization \cite{Piguet:1995er}. In Section 4 we discuss the 
anomalous dimensions of $(A^h_\mu A^h_\mu)$ and of the composite operator $A^h_\mu$ by means of the  
renormalization group equations. Section 5 contains our conclusion. A few appendices collect more details about 
the construction and the properties of the operator $(A^h_\mu A^h_\mu)$.

\section{A local framework for the operator $(A^h_\mu A^h_\mu)$}
\label{local_framework}

 Our first task will be that of finding a local framework for the non-local operator $(A^h_\mu A^h_\mu)$ 
of expression (\ref{min1}). To that purpose we start with the standard Faddeev-Popov action of Yang-Mills theory quantized in 
linear covariant gauges with the inclusion of the mass operator (\ref{min1}) as well as of a constraint enforcing the 
transversality of the field configuration $A_\mu^h$, eq.\eqref{min0}, {\it i.e.} we consider the action 
\begin{equation}
S = S_{FP}+
\int d^4x \left(\tau^{a}\,\partial_{\mu}A^{h,a}_{\mu}
+\frac{m^{2}}{2}\,A^{h,a}_{\mu}A^{h,a}_{\mu}\right)  \;, \label{act1}
\end{equation} 
where $S_{FP}$ stands for the Faddeev-Popov action in linear covariant gauges   
\begin{equation}
S_{FP} = \int d^{4}x\,\bigg(\frac{1}{4}\,F^{a}_{\mu\nu}F^{a}_{\mu\nu}+\frac{\alpha}{2}\,b^{a}b^{a}
+ib^{a}\,\partial_{\mu}A^{a}_{\mu}
+\bar{c}^{a}\,\partial_{\mu}D^{ab}_{\mu}c^{b}
\bigg)\,
\label{S_FP}
\end{equation}
and where we have introduced the operator $(A^h_\mu A^h_\mu)$ through the mass parameter $m^2$. Also, the transversality of $A_\mu^h$ is enforced by the 
Lagrange multiplier $\tau^{a}$. \\\\Since the expression for $(A^h_\mu A^h_\mu)$ given in (\ref{min1}) is an infinite sum of nonlocal
terms in the gauge field, the action \eqref{act1} should be first put in a local form 
before it can be of any practical use.
Following  \cite{Lavelle:1995ty,Ferrari:2004pd,Capri:2016aqq}, this goal can 
be achieved by the introduction of an auxiliary localizing Stueckelberg field $\xi^a$, whose role is to give, for each gauge 
field $A_\mu$, its corresponding configuration that minimizes the functional $A^2$, {\it i.e.}, 
$A^h_\mu$. This is most naturally implemented by defining a field $h$
which effectively acts on $A_\mu$ as a gauge transformation would act, in order to provide the 
minimizing configuration $A^h$, that is,
\begin{equation}\label{local_Ah}
A^{h}_{\mu} \equiv A^{h,a}_{\mu}\,T^{a}=h^{\dagger}A_{\mu}h+\frac{i}{g}h^{\dagger}\partial_{\mu}h.
\end{equation}
with
\begin{equation}
h=e^{ig\xi}=e^{ig\xi^{a}T^{a}},     \label{hxi}
\end{equation}
where $\{T^a\}$ are the generators of the gauge group $SU(N)$ and $\xi^{a}$ is a Stueckelberg field. 
\\\\Therefore, by substituting the expression (\ref{local_Ah}) for $A^h$ in the action \eqref{act1}, 
we now have a local theory in terms of the field $\xi$. The price 
one has to pay to have such a local theory is a non-polynomial action. Indeed, by expanding 
(\ref{local_Ah}), one finds an infinite series whose first terms are
\begin{equation}\label{Ah_expansion}
 (A^{h})^{a}_{\mu}=A^{a}_{\mu}-D^{ab}_{\mu}\xi^{b}-\frac{g}{2}f^{abc}\xi^{b}D^{cd}_{\mu}\xi^{d}
 +\mathcal{O}(\xi^{3})\,,   
 \end{equation}
where 
\begin{equation}
 D_\mu^{ab} = \delta^{ab}\partial_\mu - gf^{abc}A_\mu^c 
\end{equation}
is the covariant derivative in the adjoint representation. 
\\\\The nonlocal expression (\ref{min0}) for 
$A^h_\mu$ in terms of the gauge field $A_\mu$  can 
be recovered by imposing the transversality condition $\partial_\mu A^h_\mu=0$, {\it i.e.} after taking 
the divergence of both sides of (\ref{Ah_expansion}), equating it to zero and solving for the Stueckelberg 
field $\xi^a$ (see eqs.\eqref{A1},\eqref{hh2},\eqref{phi0},\eqref{minn2} of Appendix \eqref{apb}). This 
check is not only important for the consistency of the present framework but it also makes it clear that, 
due to the transversality condition enforced by the Lagrange multiplier $\tau^a$,  the Stueckelberg field 
$\xi^a$ acquires now a specific meaning:  it is precisely the field which brings a generic gauge 
configuration $A_\mu$ into the gauge-invariant and transverse field configuration $A^h_\mu$ 
which minimizes the functional $A^2_{min}$. As it will become clear in the following, this relevant feature, 
encoded in the term $\int d^4x\;\tau^{a}\,\partial_{\mu}A^{h,a}_{\mu}$, will give rise to deep differences 
between our construction and the standard Stueckelberg mass term.  The latter is known to be a 
non-renormalizable theory which has to be treated as an effective field theory  \cite{Ferrari:2004pd}. 
\\\\An important feature 
of $A^h_\mu$, as defined by eq.\eqref{local_Ah}, is its gauge invariance, that is,
\begin{equation}
A^{h}_{\mu} \rightarrow A^{h}_{\mu} \;,
\end{equation}
as can be seen from the gauge transformations  with $SU(N)$ matrix $V$
\begin{equation}
A_\mu \rightarrow V^{\dagger} A_\mu V + \frac{i}{g} V^{\dagger} \partial_ \mu V
\;, \qquad h \rightarrow V^{\dagger} h     \;, \qquad h^{\dagger} \rightarrow h^{\dagger} V \;.
\end{equation}
The local version of the action \eqref{act1}, in terms of the Stueckelberg field 
$\xi^a$, is thus given by
\begin{eqnarray}
S &=& S_{FP}+
\int d^4x \left(\tau^{a}\,\partial_{\mu}A^{h,a}_{\mu}
+\frac{m^{2}}{2}\,A^{h,a}_{\mu}A^{h,a}_{\mu}\right)\nonumber\\
&=& S_{FP}+\int d^4x \left[ \tau^a\left(A^{a}_{\mu}-D^{ab}_{\mu}\xi^{b}-\frac{g}{2}f^{abc}\xi^{b}D^{cd}_{\mu}\xi^{d}\right)\right]+
 \nonumber\\
&&+ \;\;\frac{m^2}{2}\int d^4x \left(A^{a}_{\mu}-D^{ab}_{\mu}\xi^{b}-\frac{g}{2}f^{abc}\xi^{b}D^{cd}_{\mu}\xi^{d}\right)
\left(A^{a}_{\mu}-D^{ae}_{\mu}\xi^{e}-\frac{g}{2}f^{aef}\xi^{e}D^{fg}_{\mu}\xi^{g}\right)\nonumber\\
&& \;\;\;\; + \;\;\;\cdots
\label{S_local}
\end{eqnarray}
Due to the use of the auxiliary Stueckelberg field $\xi^a$, expression \eqref{S_local} exhibits a non-polynomial character. 
At first sight, this feature might seem to jeopardize  its renormalizability. Nevertheless, this will not be the case, 
as we shall prove in the following.

Before entering into the detailed proof of the renormalizability, it is worth addressing the issue of the BRST symmetry as well 
as taking a look at the propagators of the elementary fields in order to achieve a better understanding of our action as compared 
to the usual standard massive Stueckelberg theory.

\subsection{BRST invariance}
The local action $S$, eq.\eqref{S_local}, enjoys an exact BRST symmetry:
\begin{equation}
s S = 0 \;, \label{brst_s}
\end{equation}
where the nilpotent BRST transformations are given by 
\begin{eqnarray}
sA^{a}_{\mu}&=&-D^{ab}_{\mu}c^{b}\,,\nonumber \\
sc^{a}&=&\frac{g}{2}f^{abc}c^{b}c^{c}\,, \nonumber \\
s\bar{c}^{a}&=&ib^{a}\,,\nonumber \\
sb^{a}&=&0\,, \nonumber \\
s \tau^a & = & 0\,, \nonumber \\
s^2 &=0 &\;.    \label{brst}
\end{eqnarray}
From \cite{Dragon:1996tk}, for the Stueckelberg field we have,  with $i,j$ indices associated with a generic representation,
\begin{equation}
s h^{ij} = -ig c^a (T^a)^{ik} h^{kj}  \;, \qquad s (A^h)^a_\mu = 0  \;,  \label{brstst}
\end{equation}
from which the BRST transformation of the field $\xi^a$ can be evaluated iteratively, yielding
\begin{equation}
s \xi^a=  - c^a + \frac{g}{2} f^{abc}c^b \xi^c - \frac{g^2}{12} f^{amr} f^{mpq} c^p \xi^q \xi^r + O(\xi^3)    \;.
\label{eqsxi}
\end{equation}
Let us also present a second, equivalent, way of evaluating the BRST transformation of the Stueckelberg field $\xi^a$. Owing to the dimensionless character of $\xi^a$, one starts by writing 
\begin{equation}
 s \xi^a = g^{ab}(\xi) c^b   \;, \label{gab}
 \end{equation}
 where $g^{ab}(\xi)$ stands for a generic dimensionless quantity which can be expanded in power series of  $\xi^a$. Imposing now nilpotency of the BRST operator $s$, {\it i.e.} 
 \begin{equation} 
 s^2 \xi^a = s \left( g^{ab}(\xi) c^b  \right) = 0 \;, \label{np1} 
 \end{equation}
 one gets the condition 
 \begin{equation}
 \left( \frac{\partial g^{ab}(\xi) }{\partial \xi^m} \;g^{mp}(\xi) - \frac{\partial g^{ap}(\xi) }{\partial \xi^m} \;g^{mb}(\xi)  \right) = - g^{ac}(\xi) \; (g f^{cpb})    \;. 
 \end{equation}
 The above equation can be easily solved order by order by expanding the quantity $g^{ab}(\xi)$ in power series of $\xi^a$, obtaining 
 \begin{equation}
 g^{ab}(\xi) = - \delta^{ab} + \frac{g}{2} f^{abc} \xi^c - \frac{g^2}{12} f^{acd} f^{cbe} \xi^e \xi^d + O(\xi^3)   \;, 
 \end{equation}
 which gives back precisely expression \eqref{eqsxi}. \\\\Let us end this section by checking out the explicit BRST invariance of $A^h_\mu$. To that purpose, it is better to employ a
matrix notation for the fields, {\it i.e.}
\begin{eqnarray}
sA_\mu &=& -\partial_\mu c + ig [A_\mu, c] \;, \qquad s c = -ig c c     \;, \nonumber \\
s h & =& -igch \;, \qquad sh^{\dagger}  = ig h^{\dagger} c  \;, \label{mbrst}
\end{eqnarray}
with $A_\mu = A^a_\mu T^a$, $c=c^a T^a$, $\xi=\xi^a T^a$. From expression \eqref{local_Ah} we get
\begin{eqnarray}
s A^h_\mu & = & ig h^{\dagger} c \;A_\mu h + h^{\dagger}  (-\partial_\mu c + ig [A_\mu, c]) h -ig h^{\dagger} A_\mu \;c h - h^{\dagger} c \partial_\mu h + h^{\dagger} \partial_\mu(c h) \nonumber \\
&=& igh^{\dagger} c A_\mu h -  h^{\dagger}  (\partial_\mu c ) h +ig  h^{\dagger} A_\mu \;c h - ig  h^{\dagger} c \;A_\mu h -ig h^{\dagger} A_\mu c h - h^{\dagger} c \partial_\mu h + h^{\dagger} (\partial_\mu c) h + h^{\dagger} c \partial_\mu h   \nonumber \\
&=& 0 \;. \label{sah}
\end{eqnarray}

\subsection{Comparison with the standard Stueckelberg mass term} 

Let us proceed  now by discussing the existing differences between our approach, as expressed by the 
local action $S$ of eq.\eqref{S_local}, and the usual Stueckelberg mass term.  We begin by recalling 
that the standard Stueckelberg formulation amounts to adding the mass term  $\frac{m^{2}}{2} \int d^4x 
\,A^{h,a}_{\mu}A^{h,a}_{\mu}$ to the Faddeev-Popov action, yielding thus the following action 
\begin{equation}
S_{Stueck}  = S_{FP}+
\frac{m^{2}}{2} \int d^4x 
\,A^{h,a}_{\mu}A^{h,a}_{\mu}  \;, \label{actst}
\end{equation} 
where $S_{FP}$ is the Faddeev-Popov action of the linear covariant gauges, eq.\eqref{S_FP}.  
\\\\In particular, with respect to expression \eqref{S_local}, one notices the absence, in the 
standard Stueckelberg action \eqref{actst}, of the term $ \int d^4x\;\tau^{a}\,\partial_{\mu}A^{h,a}_{\mu}$ 
enforcing the transversality condition $\partial_\mu A^h_\mu=0$. This means that the Stueckelberg mass term,  
$\frac{m^{2}}{2} \int d^4x 
\,A^{h,a}_{\mu}A^{h,a}_{\mu}$, refers to a generic gauge invariant field configuration $A^h_\mu$.  One sees 
therefore that, while in the ordinary Stueckelberg action the mass term is related to a generic gauge invariant 
configuration $A^h_\mu$, in our case, besides gauge invariance, the configuration $A^h_\mu$ is further constrained 
by the transversality condition $\partial_\mu A^h_\mu=0$. Therefore, unlike the standard Stueckelberg formulation,
our action refers to a very particular and specific mass term, which is the one obtained by mininimizing the 
operator $A^2_{min}$, as precisely expressed by the presence of the term $\int d^4x\; \tau^{a}\,\partial_{\mu}A^{h,a}_{\mu}$. 
This is a non-trivial feature of our model, which makes it deeply different from the usual Stueckelberg action \eqref{actst}. 
\\\\It is instructive  to give a look at the propagators of the Stueckelberg field $\xi^a$ which follow from both 
formulations. In the case of the standard Stueckelberg action, eq.\eqref{actst}, one obtains  
\begin{equation}
\langle \xi^a(p) \xi^b(-p) \rangle_{Stueck} = \delta^{ab} \left( 1 + \frac{\alpha m^2}{p^2}  \right) \frac{1}{m^2p^2}  \;. \label{xist} 
\end{equation}
This expression captures in a direct and simple way all drawbacks of the standard Stueckelberg formulation, as reviewed in 
\cite{Ferrari:2004pd}.  One notices, in particular, the presence of the mass parameter $m^2$ in the denominator of 
\eqref{xist}, a feature which persists even in the Landau gauge, corresponding to $\alpha=0$, namely 
\begin{equation}
\langle \xi^a(p) \xi^b(-p) \rangle_{Stueck}^{\alpha=0} = \frac{ \delta^{ab} }  {m^2p^2}  \;. \label{xistl} 
\end{equation}
As one can easily figure out, this property prevents the renormalizability of the standard Stueckelberg formulation 
\cite{Ferrari:2004pd}. In fact, due to the presence of the parameter $m^2$ in the denominator of expressions \eqref{xist}, 
\eqref{xistl}, non power-counting renormalizable divergences in the inverse of the mass $m^2$ will show up, invalidating 
the perturbative loop expansion. As discussed in \cite{Ferrari:2004pd}, the theory stemming from the action \eqref{actst} 
has to be treated within the realm of an effective non-renormalizable quantum field theory. \\\\Instead, the inclusion of 
the term  $\int d^4x\;\tau^{a}\,\partial_{\mu}A^{h,a}_{\mu}$ leads to a deep modification of the Stueckelberg propagator. 
In fact, from the quadratic part of the action $S$, eq.\eqref{S_local}, one gets (see also Appendix  \eqref{appb} where 
the complete list of propagators has been given)
\begin{equation}
\langle\xi^a(p)\xi^b(-p)\rangle_{S}=\alpha \frac{\delta^{ab}}{p^4}    \;. \label{pxiS}
\end{equation}
Expression \eqref{pxiS} displays several properties. First of all, unlike the propagator of eq.\eqref{xist},  one notices 
the absence of the mass parameter $m^2$. As far as the UV behaviour is concerned, expression \eqref{pxiS} does not pose 
any problem for the validity of the power counting, a property which will ensure in fact the all-order renormalizability 
of the model, as it will be proven in detail in the next Section. Another interesting feature displayed by expression 
\eqref{pxiS} is the decoupling nature of the Stueckelberg field in the Landau gauge, $\alpha=0$. In fact, from Appendix  
\eqref{appb}, it turns out that 
\begin{equation} 
\langle\xi^a(p)\xi^b(-p)\rangle_{S}^{\alpha=0} = \langle A^a_\mu(p) \xi^b(-p) \rangle_{S}^{\alpha=0} = \langle A_\mu^a(p) \tau^b(-p) \rangle_{S}^{\alpha} =0   \;. \label{pxiSd}
\end{equation}
This is a remarkable property of the Landau gauge, which expresses in terms of Feynman rules the decoupling of the 
Stueckelberg field $\xi^a$. It reflects the expected fact that, when $\partial_\mu A_\mu=0$, the higher order terms 
of the infinite series \eqref{min0} become harmless, due to the presence of the divergence $\partial_\mu A_\mu$. 
Equation \eqref{pxiSd} reveals in a clear way the deep difference existing between the present formulation and the 
standard Stueckelberg one for which, even in the Landau gauge, the  field $\xi^a$ does not decouple, see eq.\eqref{xistl}. 
To some extent, property  \eqref{pxiSd} makes almost immediate the perturbative renormalizability of the action $S$,  
eq.\eqref{S_local}, in the Landau gauge. \\\\Before ending this section, it is worth spending a few words on the 
possible implications of the existence of a double pole, at vanishing Euclidean momentum $p^2=0$, in the Stueckelberg 
propagator \eqref{pxiS}. Even if such a behaviour does not pose problems for the UV power-counting, 
it might give rise to unwanted infrared divergences in the explicit loop calculations. For that, a BRST 
invariant infrared regularization will be presented in the next subsection, relying on a nice property of the BRST 
transformation of the Stueckelberg field $\xi^a$. Moreover,  we underline  the presence, in expression \eqref{pxiS}, 
of the gauge parameter $\alpha$. This is a welcome feature. In fact, owing to the BRST invariance of the theory, it 
turns out that the correlation functions $\langle O(x) O(y) \rangle$ of BRST-invariant composite operators 
$O(x)$ are independent from the gauge parameter $\alpha$, see ref.\cite{Capri:2016aqq} for a recent  algebraic  proof 
of this statement. This property, combined with the aforementioned BRST-invariant infrared regularization and with the 
decoupling nature of the Stueckelberg field $\xi^a$ in the Landau gauge, ensures that the gauge invariant correlators 
$\langle O(x) O(y) \rangle$ are infrared safe. \\\\Finally, we restate the Euclidean 
nature of our construction, {\it i.e} we shall not attempt to provide a possible Minkowski interpretation for the action 
$S$, eq.\eqref{S_local}. Without entering into details, it will suffice to mention that  we expect a violation of 
perturbative unitary in Minkowski space, even if our model displays an exact BRST symmetry. This is precisely 
corroborated by the presence of a double pole in the propagator of the Stueckelberg field. Multipole fields are 
known in fact to give problems with perturbative unitarity. A nice example of this is offered by the non-local 
mass operator $F_{\mu\nu} (D^2)^{-1} F_{\mu\nu}$ which has been studied in  detail in \cite{Capri:2005dy,Capri:2006ne,Dudal:2007ch}. 
Similarly to the present case, the non-local operator $F_{\mu\nu} (D^2)^{-1} F_{\mu\nu}$ can be cast in local form 
by introducing a set of suitable auxiliary fields, so that a local formulation can be constructed at the end, 
enjoying an exact BRST symmetry \cite{Capri:2005dy,Capri:2006ne,Dudal:2007ch}. The resulting 
action turns out to be renormalizable \cite{Capri:2005dy,Capri:2006ne}. Nevertheless, it violates perturbative 
unitarity due to the presence of  multipole fields \cite{Dudal:2007ch}. We point out that the operator 
$F_{\mu\nu} (D^2)^{-1} F_{\mu\nu}$ is the first term of the infinite series of the gauge invariant expansion for the 
operator $A^2_{min}$, as one sees from eq.\eqref{zzw}. We expect thus that the same problems encountered in the 
analysis of the perturbative unitarity for the operator $F_{\mu\nu} (D^2)^{-1} F_{\mu\nu}$ will show up also in 
the case of $A^2_{min}$. \\\\Though, as it stands, the Euclidean action $S$, eq.\eqref{S_local}, turns out to be 
useful in order to study non-perturbative aspects of confining Euclidean Yang-Mills theories. In particular, 
expression \eqref{S_local}  arises within the context of the BRST-invariant formulation of the Gribov-Zwanziger 
theory  recently achieved in \cite{Capri:2015ixa,Capri:2015nzw,Capri:2016aqq}, which takes into account the 
non-perturbative effects of 
the Gribov copies. In addition, the action $S$ can be seen as the BRST-invariant extension in linear covariant 
gauges of the effective model considered by Tissier and Wschebor in the Landau gauge in order to study the positivity 
violation of the gluon propagator \cite{Tissier:2010ts,Tissier:2011ey}. Lastly, as already pointed out in the 
Introduction, the action \eqref{S_local} might enable us to investigate the formation of the dimension two-condensate 
$\langle A^h_\mu A^h_\mu \rangle$ in a BRST-invariant and $\alpha$-independent way.

\subsection{Infrared BRST-invariant regularization for the Stueckelberg field $\xi$} \label{subsection:StueckelbergMass}

As mentioned before, the propagator for the Stueckelberg field in expression \eqref{pxiS}  could give rise to potential IR divergences when performing explicit loop calculations. Though, as outlined in \cite{Capri:2016aqq}, it turns out to be possible to introduce an IR regularizing mass term for the Stueckelberg field compatible with the BRST invariance.  For the benefit of the reader, let us reproduce here the construction of \cite{Capri:2016aqq}. It relies on a nice property displayed by the BRST transformation of the field $\xi^a$  given in eqs.\eqref{brstst}, \eqref{eqsxi}, namely  
\begin{equation}
s\left(\frac{\xi^a\xi^a}{2}\right)=-\xi^a c^a \,,    \label{prost}
\end{equation}
as it follows from  eq.\eqref{mbrst}, {\it  i.e.}
\begin{equation}
 s (e^{ig\xi}) = - ig c e^{ig\xi} \;.  \label{ig1}
\end{equation}
Expanding the exponential in Taylor series, one gets
\begin{equation}
 s \left( 1 + ig \xi - \frac{g^2}{2} \xi \xi - i \frac{g^3}{3!} \xi \xi \xi + \cdot  \cdot  \right)  = -igc \left(   1 + ig \xi - \frac{g^2}{2} \xi \xi - i \frac{g^3}{3!} \xi \xi \xi + \cdot  \cdot \right)  \;. \label{ig2}
\end{equation}
Multiplying both sides of eq.\eqref{ig2} by $\xi$, yields
\begin{equation}
\xi \; s \left( 1 + ig \xi - \frac{g^2}{2} \xi \xi - i \frac{g^3}{3!} \xi \xi \xi + \cdot  \cdot  \right)  = -ig \xi \;c \left(   1 + ig \xi - \frac{g^2}{2} \xi \xi - i \frac{g^3}{3!} \xi \xi \xi + \cdot  \cdot \right)  \;. \label{ig3}
\end{equation}
Equating now order by order in $g$ the expression \eqref{ig3} immediately provides eq.\eqref{prost}.

\noindent Due to equation \eqref{prost}, we can introduce the following BRST-exact term
\begin{equation}
 S_{IRR}=\int d^4x \frac{1}{2} s\left(\rho\xi^a\xi^a\right) =\int d^4x \left( \frac{1}{2}M^4\xi^a\xi^a+\rho\xi^ac^a  \right) \,,
\label{SIRR}
\end{equation}
where $(\rho, M)$ are constant parameters transforming as
 \begin{equation}
 s\rho=M^{4}\,,\qquad sM^{4}=0\,.
 \end{equation}
 As it is apparent, the action $(S+S_{IRR})$ is BRST invariant, {\it i.e.} 
 \begin{equation}
 s (S+S_{IRR}) =0  \;. \label{invna}
 \end{equation}
The parameter $\rho$ has ghost number $-1$, while $M$ has ghost number zero. From equation \eqref{SIRR}, it turns out that the propagator of the Stueckelberg field  $\xi^a$ behaves now like
 \begin{equation}
 \langle \xi^a(p) \xi^b(-p) \rangle_{S+S_{IRR}} = \delta^{ab}  \frac{\alpha}{p^4+\alpha M^4} \;, \label{regprop}
 \end{equation} 
 showing that the mass parameter $M$ introduces an IR regularization in a BRST-invariant way. In Appendix  \eqref{appb} one finds the whole list of all propagators of the elementary fields evaluated in  presence of the parameters $(\rho, M)$, which have to be set to zero  at the very end of the computation of the correlation functions.

\section{Renormalizability} \label{Renormalizability} 
We are now ready to face the issue of the all order renormalizability  of the action $S$,  eq.\eqref{S_local}. For later convenience, it turns out to be helpful to employ a slightly different parametrization, redefining the gauge parameter $\alpha$ as well as  the  gauge,  Lagrange multiplier, and Stueckelberg fields as 
\begin{equation}
A^{a}_{\mu}\to  \frac{1}{g}\,A^{a}_{\mu}\,,\qquad
	b^{a}\to gb^{a}\,, \qquad \xi^a \to \frac{1}{g} \xi^a\;, \qquad \alpha\to\frac{\alpha}{g^{2}} \;. \label{red}   
\end{equation}
Accordingly, for the field strength and the covariant derivative, we get 
\begin{eqnarray}
F^{a}_{\mu\nu}&=&\partial_{\mu}A^{a}_{\nu}-\partial_{\nu}A^{a}_{\mu}+f^{abc}A^{b}_{\mu}A^{c}_{\nu}\,,\label{F_new}\\
D^{ab}_{\mu}&=&\delta^{ab}\partial_{\mu}-f^{abc}A^{c}_{\mu}\,,\label{D_new}
\end{eqnarray}
while for the action $S$ 
\begin{eqnarray}
S &=& S_{FP} + \int d^4x \left(\tau^{a}\,\partial_{\mu}A^{h,a}_{\mu} +\frac{m^{2}}{2}\,A^{h,a}_{\mu}A^{h,a}_{\mu}\right)\;, \label{actnp}  \\
S_{FP} &=&  \int d^{4}x\,\bigg(\frac{1}{4g^2}\,F^{a}_{\mu\nu}F^{a}_{\mu\nu}+\frac{\alpha}{2}\,b^{a}b^{a}
+ib^{a}\,\partial_{\mu}A^{a}_{\mu}
+\bar{c}^{a}\,\partial_{\mu}D^{ab}_{\mu}c^{b}
\bigg)
  \label{nact} 
\end{eqnarray} 
where 
\begin{equation}\label{nlocal_Ah}
A^{h}_{\mu} \equiv A^{h,a}_{\mu}\,T^{a}=h^{\dagger}A_{\mu}h+i h^{\dagger}\partial_{\mu}h \;, \qquad 
h=e^{i\xi^{a}T^{a}} \;. 
\end{equation}
Also, for the BRST transformation, we have 
\begin{eqnarray}
sA^{a}_{\mu}&=&-D^{ab}_{\mu}c^{b}\,,\nonumber\\
sc^{a}&=&\frac{1}{2}f^{abc}c^{b}c^{c}\,,\nonumber\\
s\bar{c}^{a}&=&ib^{a}\,,\nonumber\\
sb^{a}&=&0\,, \nonumber \\
s\tau^a & = & 0\;,  \label{nbrst}
\end{eqnarray}
and 
\begin{equation}
s\xi^{a}=  g^{ab}(\xi)c^{b}\,,\qquad  
 g^{ab}(\xi) = - \delta^{ab} + \frac{1}{2} f^{abc} \xi^c - \frac{1}{12} f^{acd} f^{cbe} \xi^e \xi^d + O(\xi^3)   \;,  \label{nxibrst}
 \end{equation}
 with
\begin{equation}
s S =0 \;. \label{nsinv}
\end{equation}
The usefulness of the new parametrization in eqs.\eqref{actnp},\eqref{nact} relies on the property that, acting on the action S with the differential operator ${g^2} \frac{\partial }{\partial g^2}$, gives directly the gauge invariant quantity $\int d^4x F^a_{\mu\nu} F^a_{\mu\nu}$, {\it i.e.} 
\begin{equation}
{g^2} \frac{\partial S}{\partial g^2} = - \frac{1}{4g^2} \int d^{4}x \,F^{a}_{\mu\nu}F^{a}_{\mu\nu}  \;, 
\end{equation} 
a feature which will be helpful in order to write down the parametric form of the most general counterterm allowed by the quantum corrections. 
\\\\Let us proceed by identifying the Ward identities of the model. To that purpose, following the algebraic renormalization set 
up \cite{Piguet:1995er}, we introduce a set of BRST-invariant external sources 
$(J(x), \mathcal{J}(x)^{a}_{\mu}, \Omega^a_\mu(x), L^a(x), K^a(x))$ coupled to the composite operators 
$(A^h_\mu(x) A^h_\mu(x))$ and $A^h_\mu(x)$ as well as to the non-linear BRST variation of the fields $(A^a_\mu, c^a, \xi^a)$, 
namely we consider the classical complete BRST-invariant action $\Sigma$ defined by 
\begin{eqnarray}
\Sigma&=&S_{FP}+\int d^{4}x\bigg(\tau^{a}\,\partial_{\mu}A^{h,a}_{\mu}
+\frac{J}{2}\,A^{h,a}_{\mu}A^{h,a}_{\mu}
+\mathcal{J}^{a}_{\mu}A^{h,a}_{\mu}
-\Omega^{a}_{\mu}\,D^{ab}_{\mu}c^{b}
+\frac{1}{2}f^{abc}L^{a}c^{b}c^{c}\nonumber\\
&&+K^{a}\,g^{ab}(\xi)c^{b}
+\frac{\zeta}{2}\,J^{2}\bigg)\,,
\label{The_action}
\end{eqnarray}
with
\begin{equation}
sJ=s\mathcal{J}^{a}_{\mu}=s\Omega^{a}_{\mu}=sL^{a}=sK^{a}=0\,, \label{insc}
\end{equation}
which ensure the BRST invariance of $\Sigma$
\begin{equation}
s \Sigma =0 \;. \label{invS}
\end{equation}
The action $S$, eq.\eqref{actnp}, can be recovered from $\Sigma$, modulo a constant vacuum term $V \frac{\zeta}{2}m^4$,  by setting the sources $(J, \mathcal{J}^{a}_{\mu}, \Omega^a_\mu, L^a, K^a)$ equal to
\begin{equation}
J\Big|_{phys} = m^2 \;, \qquad \mathcal{J}^{a}_{\mu}\Big|_{phys}=\Omega^{a}_{\mu}\Big|_{phys}=L^{a}\Big|_{phys}=K^{a}\Big|_{phys}=0\,,
\end{equation}  
{\it i.e.} 
\begin{equation}
 \Sigma \Big|_{phys}= S +  V \frac{\zeta}{2}m^4  \;, \label{phys}
\end{equation}
where $V$ stands for the Euclidean space-time volume. The parameter $\zeta$ is a dimensionless free parameter which enables us to take into account possible divergences affecting the vacuum term $J^{2}(x)$  \cite{Verschelde:2001ia,Dudal:2003vv,Browne:2003uv},  allowed by power-counting due to the fact that source $J(x)$ has dimension two. Let us also mention that the vacuum term $\frac{\zeta}{2}\,J^{2}$ is required in order to investigate the formation of the dimension two  condensate $\langle A^h_\mu(x) A^h_\mu(x) \rangle$ via evaluation of the corresponding effective potential, see \cite{Verschelde:2001ia,Dudal:2003vv,Browne:2003uv}. In particular, the parameter $\zeta$ can be made a function of the coupling constant $g$ in such a way that the generating functional of the correlation functions of the theory obeys a homogeneous renormalization group equation \cite{Verschelde:2001ia,Dudal:2003vv,Browne:2003uv}, a result which will be employed in Section \eqref{andim} in order to determine the anomalous 
dimensions of the operators $(A^h_\mu A^h_\mu)$ and $A^h_\mu$.

\subsection{Ward identities}

The BRST symmetry stated in the previous section can be immediately  written as a functional identity. The complete classical action $\Sigma$ turns out to fulfill the following Ward identities: 
\begin{itemize} 

\item{The Slavnov-Taylor identity}
\begin{eqnarray}
\mathcal{S}(\Sigma)&\equiv&\int d^{4}x\,\bigg(
\frac{\delta\Sigma}{\delta\Omega^{a}_{\mu}}\frac{\delta\Sigma}{\delta A^{a}_{\mu}}
+\frac{\delta\Sigma}{\delta L^{a}}\frac{\delta\Sigma}{\delta c^{a}}
+\frac{\delta\Sigma}{\delta K^{a}}\frac{\delta\Sigma}{\delta\xi^{a}}
+ib^{a}\frac{\delta\Sigma}{\delta\bar{c}^{a}}\bigg)=0\,.    \label{sti}
\end{eqnarray}
In view of the algebraic characterization of the counterterm, we introduce the so-called linearized Slavnov-Taylor operator  $\mathcal{B}_{\Sigma}$ \cite{Piguet:1995er} defined as 
\begin{equation}
\mathcal{B}_{\Sigma}\equiv\int d^{4}x\,\biggl(
\frac{\delta\Sigma}{\delta\Omega^{a}_{\mu}}\frac{\delta}{\delta A^{a}_{\mu}}
+\frac{\delta\Sigma}{\delta A^{a}_{\mu}}\frac{\delta}{\delta\Omega^{a}_{\mu}}
+\frac{\delta\Sigma}{\delta L^{a}}\frac{\delta}{\delta c^{a}}
+\frac{\delta\Sigma}{\delta c^{a}}\frac{\delta}{\delta L^{a}}
+\frac{\delta\Sigma}{\delta K^{a}}\frac{\delta}{\delta\xi^{a}}
+\frac{\delta\Sigma}{\delta\xi^{a}}\frac{\delta}{\delta K^{a}}
+ib^{a}\frac{\delta}{\delta\bar{c}^{a}}\bigg)\,,
\label{ST_linearized}
\end{equation}
which, as the BRST operator $s$, turns out to be nilpotent 
\begin{equation}
\mathcal{B}_{\Sigma}\mathcal{B}_{\Sigma}=0\,.
\end{equation}

\item{The gauge-fixing condition and the anti-ghost equation \cite{Piguet:1995er} }
\begin{equation}
\frac{\delta\Sigma}{\delta b^{a}}=i\partial_{\mu}A^{a}_{\mu}+\alpha\,b^{a}\,,
\label{gauge_fixing}
\end{equation}
\begin{equation}
\frac{\delta\Sigma}{\delta\bar{c}^{a}}+\partial_{\mu}\frac{\delta\Sigma}{\delta\Omega^{a}_{\mu}}=0\,,
\label{aghost_eq}
\end{equation}
In particular, the identity \eqref{aghost_eq} ensures that the anti-ghost field $\bar{c}^{a}$ and the source $\Omega^{a}_{\mu}$ enter only through the combination
\begin{equation}
\widehat{\Omega}^{a}_{\mu}=\Omega^{a}_{\mu}+\partial_{\mu}\bar{c}^{a}\,.
\label{omega_hat}
\end{equation}
\item{The $\tau$ Ward identity}
\begin{equation}
\frac{\delta\Sigma}{\delta\tau^{a}}-\partial_{\mu}\frac{\delta\Sigma}{\delta \mathcal{J}^{a}_{\mu}}=0\,,\label{tau_ward}
\end{equation}
implying that  the field $\tau^{a}$ and the source $\mathcal{J}^{a}_{\mu}$ appear only in the combination
\begin{equation}
\widehat{\mathcal{J}}^{a}_{\mu}=\mathcal{J}^{a}_{\mu}-\partial_{\mu}\tau^{a}\,.
\label{J_hat}
\end{equation}

\end{itemize} 

\subsection{Characterization of the most general counterterm}
In order to characterize the most general invariant counterterm  which can be freely added to all orders
in perturbation theory we follow the set up of the algebraic renormalization  \cite{Piguet:1995er} and 
perturb the classical action $\Sigma$ by adding an integrated local quantity in the fields and sources, 
$\Sigma^{ct}$, with dimension bounded by four and vanishing ghost number. We demand thus that the 
perturbed action, $(\Sigma +\varepsilon\Sigma^{ct})$, where $\varepsilon$ is an expansion parameter, 
fulfills, to the first order in $\varepsilon$, the same Ward identities obeyed by the classical action 
$\Sigma$, {\it i.e.} equations \eqref{sti}, \eqref{gauge_fixing}, \eqref{aghost_eq}, 
and \eqref{tau_ward}.  
This requirement gives rise to the set of equations
\begin{eqnarray}
\mathcal{S}(\Sigma+\varepsilon\,\Sigma^{ct})&=&O(\varepsilon^{2})\,,\nonumber\\
\frac{\delta}{\delta b^{a}}(\Sigma+\varepsilon\,\Sigma^{ct})&=&i\partial_{\mu}A^{a}_{\mu}+\alpha\,b^{a}+O(\varepsilon^{2})\,,\nonumber\\
\left(\frac{\delta}{\delta\bar{c}^{a}}+\partial_{\mu}\frac{\delta}{\delta\Omega^{a}_{\mu}}\right)(\Sigma+\varepsilon\,\Sigma^{ct})&=&O(\varepsilon^{2})\,,\nonumber\\
\left(\frac{\delta}{\delta\tau^{a}}-\partial_{\mu}\frac{\delta}{\delta\mathcal{J}^{a}_{\mu}}\right)(\Sigma+\varepsilon\,\Sigma^{ct})&=&O(\varepsilon^{2})\,, 
\end{eqnarray}
yielding the following constraints on  $\Sigma^{ct}$:
\begin{eqnarray}
\mathcal{B}_{\Sigma}\Sigma^{ct}&=&0\,,\label{B_constraint}\\
\frac{\delta}{\delta b^{a}}\Sigma^{ct}&=&0\,,\label{b_constraint}\\
\left(\frac{\delta}{\delta\bar{c}^{a}}+\partial_{\mu}\frac{\delta}{\delta\Omega^{a}_{\mu}}\right)\Sigma^{ct}&=&0\,,\label{cbar_constraint}\\
\left(\frac{\delta}{\delta\tau^{a}}-\partial_{\mu}\frac{\delta}{\delta\mathcal{J}^{a}_{\mu}}\right)\Sigma^{ct}&=&0\,.\label{tau_constraint}
\end{eqnarray}

\noindent From the constraint \eqref{b_constraint} it follows that $\Sigma^{ct}$ is independent from the Lagarange multiplier  
$b^{a}$, while equations \eqref{cbar_constraint} and \eqref{tau_constraint}  ensure that $\Sigma^{ct}$ depends only on the 
combinations $\widehat{\Omega}^{a}_{\mu}=\Omega^{a}_{\mu}+\partial_{\mu}\bar{c}^{a}$ and 
$\widehat{\mathcal{J}}^{a}_{\mu}=\mathcal{J}^{a}_{\mu}-\partial_{\mu}\tau^{a}$ of equations \eqref{omega_hat},\eqref{J_hat}. 
\\\\From equation \eqref{B_constraint} one learns that $\Sigma^{ct}$ belongs to the cohomolgy \cite{Piguet:1995er} of the 
linearized Slavnov-Taylor operator $\mathcal{B}_{\Sigma}$ in the space of the integrated local quantities in the fields 
and sources of dimension 4 and ghost number zero. Therefore, we can set 
\begin{equation}
\Sigma^{ct}=\Delta+\mathcal{B}_{\Sigma}\Delta^{(-1)}\,,
\label{cohom}
\end{equation} 
where $\Delta^{(-1)}$ denotes a four-dimensional integrated quantity in the fields and sources with ghost number $-1$. 
The term $\mathcal{B}_{\Sigma}\Delta^{(-1)}$ in equation \eqref{cohom} corresponds to the trivial solution, {\it i.e.} 
to the exact part of the cohomology of $\mathcal{B}_{\Sigma}$. Instead, the quantity  $\Delta$ identifies the non-trivial 
solution, {\it i.e.} the cohomology of $\mathcal{B}_{\Sigma}$, meaning that  $\Delta \neq \mathcal{B}_{\Sigma} Q$, for 
some local integrated  $Q$. \\\\From the general results on the cohomology of Yang-Mills theories \cite{Piguet:1995er}, 
and with the help of Table~\ref{table}, where the dimension and the ghost number of all fields and sources are displayed, 
it follows that $\Delta$ and $\Delta^{(-1)}$ can be written as
\begin{table}
\centering
\begin{tabular}{|c|c|c|c|c|c|c|c|c|c|c|c|}
\hline
Fields \& sources &$A$&$b$&$c$&$\bar{c}$&$\xi$&$\tau$&$\Omega$&$L$&$K$&$J$&$\mathcal{J}$\\
\hline\hline
Dimension &1&2&0&2&0&2&3&4&4&2&3\\
\hline
Ghost number&0&0&1&$-1$&0&0&$-1$&$-2$&$-1$&0&0\\
\hline
\end{tabular}
\caption{The quantum numbers of fields and sources.}
\label{table}
\end{table}
\begin{eqnarray}
\Delta&=&\int d^{4}x\,\bigg[
\frac{c_0}{4g^{2}}\,F^{a}_{\mu\nu}F^{a}_{\mu\nu}
+c_{1}\,(\partial_{\mu}A^{h,a}_{\mu})\,\partial_{\nu}A^{h,a}_{\nu}
+c_{2}\,(\partial_{\mu}A^{h,a}_{\nu})\,\partial_{\mu}A^{h,a}_{\nu}
+c_{3}\,f^{abc}A^{h,a}_{\mu}A^{h,b}_{\nu}\,\partial_{\mu}A^{h,c}_{\nu}\nonumber\\
&&
+\lambda^{abcd}\,A^{h,a}_{\mu}A^{h,b}_{\mu}A^{h,c}_{\nu}A^{h,d}_{\nu}
+\widehat{\mathcal{J}}^{a}_{\mu}\,\mathcal{O}^{a}_{\mu}(A,\xi)
+J\,\mathcal{O}(A,\xi)
+c_{4}\,\frac{\zeta}{2}\,J^{2}
\bigg]\,,
\label{Delta}
\end{eqnarray}
and
\begin{eqnarray}
\Delta^{(-1)}&=&\int d^{4}x\,\bigg[
f^{ab}_{1}(\xi)\,\widehat\Omega^{a}_{\mu}A^{b}_{\mu}
+f^{ab}_{2}(\xi)\,L^{a}c^{b}
+f^{ab}_{3}(\xi)\,K^{a}\xi^{b}\bigg]\,,
\label{Delta_minus_one}
\end{eqnarray}
where we have taken into account the gauge invariant nature of the field $A^h_\mu$, {\it i.e.} 
\begin{equation}
\mathcal{B}_{\Sigma} A^{h,a}_\mu = s A^{h,a}_\mu = 0 \;, \qquad   A^{h,a}_\mu \neq \mathcal{B}_{\Sigma}  (\hat \rho^a_\mu)  \;, \label{sahinv}
\end{equation}
for any $\hat\rho_\mu^a$.
The parameters $(c_0, c_{1}, c_{2}, c_{3}, c_{4}, \lambda^{abcd})$ in expression \eqref{Delta} are free dimensionless coefficients, while $\mathcal{O}^{a}_{\mu}(A,\xi)$ and $\mathcal{O}(A,\xi)$ stand for generic local quantities with dimension 1 and 2 and ghost number zero, respectively, depending only on the fields $A^{a}_{\mu}$ and $\xi^{a}$. Also, the quantities $f_{1}^{ab}(\xi)$, $f_{2}^{ab}(\xi)$ and $f_{3}^{ab}(\xi)$ in expression \eqref{Delta_minus_one} are arbitrary power series in $\xi$ with ghost number zero,  allowed by the dimensionless character of the Stueckelberg field $\xi^a$. \\\\Imposing now the constraint \eqref{B_constraint}, one immediately gets 
\begin{eqnarray}
\mathcal{B}_{\Sigma} \mathcal{O}^{a}_{\mu}(A,\xi) & = & s \mathcal{O}^{a}_{\mu}(A,\xi) = 0 \;, \label{o1} \\
\mathcal{B}_{\Sigma} \mathcal{O}(A,\xi) & = & s \mathcal{O}(A,\xi) =0 \;, \label{o2}
\end{eqnarray}
meaning that $\mathcal{O}^{a}_{\mu}(A,\xi)$ and $\mathcal{O}(A,\xi)$ have to be BRST invariant. Let us work out in detail the most general solutions of eqs.\eqref{o1},\eqref{o2}, beginning with eq.\eqref{o1}. Taking into account that the operator $\mathcal{O}^{a}_{\mu}(A,\xi)$ has dimension 1, ghost number 0, and carries both  color and  Lorentz indices, it  can be parametrized as
\begin{equation}
\mathcal{O}^{a}_{\mu}(A,\xi)=\sigma^{ab}(\xi)\,A^{b}_{\mu}+\omega^{ab}(\xi)\,\partial_{\mu}\xi^{b}\,, \label{p1}
\end{equation} 
where $(\sigma^{ab}(\xi), \omega^{ab}(\xi))$ are dimensionless quantities in the Stueckelberg field $\xi^a$. Making use of expression \eqref{nlocal_Ah}, it turns out to be useful to replace $A^{a}_{\mu}$ by the gauge invariant field $A^{h,a}_{\mu}$, upon a redefinition of the quantities $(\sigma^{ab}(\xi), \omega^{ab}(\xi))$, {\it i.e.}
\begin{equation}
\mathcal{O}^{a}_{\mu}(A,\xi)=\hat\sigma^{ab}(\xi)\,A^{h,b}_{\mu}+\hat\omega^{ab}(\xi)\,\partial_{\mu}\xi^{b}\,.
\end{equation} 
Therefore, from condition \eqref{o1} one gets 
\begin{equation}
\frac{\partial\hat\sigma^{ab}}{\partial\xi^{c}}\,g^{cd}c^{d}\,A^{h,b}_{\mu}
+\left(\frac{\partial\hat\omega^{ad}}{\partial\xi^{b}}g^{bc}
+\hat\omega^{ab}\,\frac{\partial g^{bc}}{\partial\xi^{d}}\right)\,c^{c}\,\partial_{\mu}\xi^{d}
+\hat\omega^{ab}g^{bc}\,\partial_{\mu}c^{c}=0\,,
\end{equation}
which immediately gives 
\begin{eqnarray}
\frac{\partial\hat\sigma^{ab}}{\partial\xi^{c}} &=&0  \Rightarrow \hat\sigma^{ab}=b_{1} \delta^{ab} \nonumber \;,\\
 \hat\omega^{ab}&= &0\,,
\end{eqnarray}
where $b_1$ is a constant. We conclude thus that the most general form for $\mathcal{O}^{a}_{\mu}$ is given by 
\begin{equation}
\mathcal{O}^{a}_{\mu}(A,\xi) = b_1 \,A^{h,a}_{\mu}\,.     \label{Omu}
\end{equation}
The same reasoning applies as well to the case of the operator  $\mathcal{O}(A,\xi)$ in eq.\eqref{o2}. Taking into account now that $\mathcal{O}(A,\xi)$ is of dimension two, we write 
\begin{equation}
\mathcal{O}(A,\xi)=\sigma^{ab}(\xi)\,A^{a}_{\mu}A^{b}_{\mu}+\omega^{a}(\xi)\,\partial_{\mu}A^{a}_{\mu}
+\lambda^{ab}(\xi)\,A^{a}_{\mu}\,\partial_{\mu}\xi^{b}
+\frac{\rho^{ab}(\xi)}{2}\,(\partial_{\mu}\xi^{a})\partial_{\mu}\xi^{b}
+\beta^{a}(\xi)\,\partial^{2}\xi^{a}\,,
\end{equation}
where $(\sigma^{ab}(\xi), \omega^{a}(\xi), \lambda^{ab}(\xi), \rho^{ab}(\xi), \beta^{a}(\xi))$ are dimensionless power series in $\xi$. 
Again, employing the gauge invariant variable $A^h_\mu$, we obtain, upon a redefinition of $(\sigma^{ab}, \omega^{a}, \lambda^{ab}, \rho^{ab}, \beta^{a})$, 
\begin{equation}
\mathcal{O}(A,\xi)=\hat\sigma^{ab}(\xi)\,A^{h,a}_{\mu}A^{h,b}_{\mu}+\hat\omega^{a}(\xi)\,\partial_{\mu}A^{h,a}_{\mu}
+\hat\lambda^{ab}(\xi)\,A^{h,a}_{\mu}\,\partial_{\mu}\xi^{b}
+\frac{\hat\rho^{ab}(\xi)}{2}\,(\partial_{\mu}\xi^{a})\partial_{\mu}\xi^{b}
+\hat\beta^{a}(\xi)\,\partial^{2}\xi^{a}\,.
\end{equation}
From equation \eqref{o2} we have 
\begin{eqnarray}
0&=&\frac{\partial\hat\sigma^{ab}}{\partial\xi^{d}}g^{dc}\,\,A^{h,a}_{\mu}A^{h,b}_{\mu}c^{c}
+\frac{\partial\hat\omega^{a}}{\partial\xi^{c}}g^{cb}\,\,(\partial_{\mu}A^{h,a}_{\mu})c^{b}
+\left(\frac{\partial\hat\lambda^{ab}}{\partial\xi^{d}}g^{dc}+\hat\lambda^{ad}\frac{\partial g^{dc}}{\partial\xi^{b}}\right)\,A^{h,a}_{\mu}(\partial_{\mu}\xi^{b})c^{c}\nonumber\\
&&+\hat\lambda^{ac}g^{cb}\,\,A^{h,a}_{\mu}\,\partial_{\mu}c^{b}
+\left(\frac{1}{2}\frac{\partial\hat\rho^{ab}}{\partial\xi^{d}}g^{dc}
+\hat\rho^{ad}\frac{\partial g^{dc}}{\partial\xi^{b}}
+\hat\beta^{d}\frac{\partial^{2}g^{dc}}{\partial\xi^{a}\partial\xi^{b}}\right)\,(\partial_{\mu}\xi^{a})
(\partial_{\mu}\xi^{b})c^{c}\nonumber\\
&&+\left(\hat\rho^{ac}g^{cb}+2\hat\beta^{c}\frac{\partial g^{cb}}{\partial\xi^{a}}\right)\,(\partial_{\mu}\xi^{a})\partial_{\mu}c^{b}
+\left(\frac{\partial\hat\beta^{a}}{\partial\xi^{c}}g^{cb}+\hat\beta^{c}\frac{\partial g^{cb}}{\partial\xi^{a}}\right)\,(\partial^{2}\xi^{a})c^{b}\nonumber\\
&&+\hat\beta^{b}g^{ba}\,\,\partial^{2}c^{a}\,,
\end{eqnarray}
from which it follows that 
\begin{eqnarray}
\hat\lambda^{ab} &=& \hat\rho^{ab}=\hat\beta^{a}=0\,, \nonumber \\
\frac{\partial\hat\sigma^{ab}}{\partial\xi^{d}}&=& 0 \Rightarrow \hat\sigma^{ab} = \frac{b_2}{2} \delta^{ab} \;, \nonumber \\
\frac{\partial\hat\omega^{a}}{\partial\xi^{b}}& =& 0 \Rightarrow \hat\omega^{a}=0\,\,\mbox{(by color invariance)} \;,
\end{eqnarray}
where $b_2$ is a free coefficient. Finally, for the operator $\mathcal{O}(A,\xi)$, we have
\begin{equation}
\mathcal{O}(A,\xi)= \frac{b_2}{2} \,A^{h,a}_{\mu}A^{h,a}_{\mu}\,.
\end{equation}
Therefore, for the most general counterterm, eq.\eqref{cohom}, we get 
\begin{eqnarray}
\Delta&=&\int d^{4}x\,\bigg[
\frac{c_0}{4g^{2}}\,F^{a}_{\mu\nu}F^{a}_{\mu\nu}
+c_{1}\,(\partial_{\mu}A^{h,a}_{\mu})\,\partial_{\nu}A^{h,a}_{\nu}
+c_{2}\,(\partial_{\mu}A^{h,a}_{\nu})\,\partial_{\mu}A^{h,a}_{\nu}
+c_{3}\,f^{abc}A^{h,a}_{\mu}A^{h,b}_{\nu}\,\partial_{\mu}A^{h,c}_{\nu}\nonumber\\
&&
+\lambda^{abcd}\,A^{h,a}_{\mu}A^{h,b}_{\mu}A^{h,c}_{\nu}A^{h,d}_{\nu}
+b_1 \widehat{\mathcal{J}}^{a}_{\mu}\, A^{h,a}_\mu 
+\frac{b_2}{2} J\, A^{h,a}_\mu A^{h,a}_\mu
+c_{4}\,\frac{\zeta}{2}\,J^{2}
\bigg]\,,
\label{Delta1}
\end{eqnarray}
and  $\Delta^{(-1)}$ is given by eq. \eqref{Delta_minus_one}.

It remains now to characterize the coefficients $(c_1, c_2, c_3, \lambda^{abcd})$. To that aim, we rely on an important property of the action $S$ in equation \eqref{actnp}. When the mass parameter $m^2$ is set to zero, {\it i.e.} $m^2=0$, expression reduces to 
\begin{eqnarray}
S_{m^2=0} &=& S_{FP} + \int d^4x \left(\tau^{a}\,\partial_{\mu}A^{h,a}_{\mu} \right)\;, \label{mactnp}  \\
S_{FP} &=&  \int d^{4}x\,\bigg(\frac{1}{4g^2}\,F^{a}_{\mu\nu}F^{a}_{\mu\nu}+\frac{\alpha}{2}\,b^{a}b^{a}
+ib^{a}\,\partial_{\mu}A^{a}_{\mu}
+\bar{c}^{a}\,\partial_{\mu}D^{ab}_{\mu}c^{b}
\bigg)  \;,
  \label{nacta1} 
\end{eqnarray} 
which coincides, modulo the term $\int d^4x \;\tau^{a}\,\partial_{\mu}A^{h,a}_{\mu}$, with the Faddeev-Popov action $S_{FP}$ of the linear covariant gauges. \\\\Nevertheless, as shown in detail in Appendix \eqref{review},  the additional term $\int d^4x\;  \tau^{a}\,\partial_{\mu}A^{h,a}_{\mu} $ has no consequences on the evaluation of the Green functions of the elementary fields $(A_\mu, b, c, \bar c)$, meaning that the correlation functions 
$\langle A_{\mu_1}(x_1) ...... A_{\mu_n}(x_n) \rangle_{S_{m^2=0}} $ evaluated with the action $S_{m^2=0}$ coincide with those computed with the Faddeev-Popov action 
$S_{FP}$, namely 
\begin{equation}
\langle A_{\mu_1}(x_1) ...... A_{\mu_n}(x_n) \rangle_{S_{m^2=0}}  = \langle A_{\mu_1}(x_1) ...... A_{\mu_n}(x_n) \rangle_{S_{FP}}   \;.  \label{statp}
\end{equation}
From this property, it follows that when the external fields  $(J,\mathcal{J},K)$ are set to zero, {\it i.e.} $(J,\mathcal{J},K)\to0$, the counterterm \eqref{Delta1},  \eqref{Delta_minus_one} has to reduce to that of the Faddeev-Popov action  in presence of the term $\int d^4x \;\tau^{a}\,\partial_{\mu}A^{h,a}_{\mu}$, namely to expressions \eqref{countact1},\eqref{pffp},\eqref{fpct} of Appendix \eqref{review}. This requirement gives 
\begin{equation}
c_0=a_0\,,\qquad c_1=c_2=c_3=0\,,\qquad \lambda^{abcd}=0\,,\qquad f^{ab}_{1}=a_1\,\delta^{ab}\,,\qquad f^{ab}_{2}=a_2\,\delta^{ab}\,.
\end{equation}  
so that for the counterterm $\Sigma^{ct}$ we obtain
\begin{eqnarray}
\Sigma^{ct}&=&\int d^{4}x\,\bigg[\,
\frac{a_0}{4g^{2}}\,F^{a}_{\mu\nu}F^{a}_{\mu\nu}
+b_{1}\,\widehat{\mathcal{J}}^{a}_{\mu}\,A^{h,a}_{\mu}
+\frac{b_{2}}{2}\,J\,A^{h,a}_{\mu}A^{h,a}_{\mu}
+b_{3}\,\frac{\zeta}{2}\,J^{2}\,
\bigg]\nonumber\\
&&+\mathcal{B}_{\Sigma}\int d^{4}x\,\bigg[\,
a_{1}\,\widehat\Omega^{a}_{\mu}A^{a}_{\mu}
+a_{2}\,L^{a}c^{a}
+K^{a}f^{a}(\xi)\,\bigg]\,,
\label{count}
\end{eqnarray}
where we have performed the following redefinitions: 
\begin{equation}
f^{a}(\xi)\equiv f_{3}^{ab}(\xi)\xi^{b}\,,\qquad b_{3}\equiv c_{4}\,.
\label{redef}
\end{equation}

\subsection{Parametric form of the counterterm and renormalization factors}
Having determined the most general form of the invariant counterterm, eq.\eqref{count}, it remains to check if $\Sigma^{ct}$ can be reabsorbed in the starting action $\Sigma$ through a redefinition of parameters, fields and sources. To that end, let us proceed by casting expression \eqref{count} in the so called parametric form.  From the expressions of the linearized Slavnov-Taylor operator $\mathcal{B}_{\Sigma}$, eq.~\eqref{ST_linearized}, we can rewrite the counterterm $\Sigma^{ct}$ as
\begin{eqnarray}
\Sigma^{ct}&=&\int d^{4}x\,\bigg(\frac{a_{0}}{4g^{2}}\,F^{a}_{\mu\nu}F^{a}_{\mu\nu}
+b_{1}\,\mathcal{J}^{a}_{\mu}A^{h,a}_{\mu}
+b_{1}\,\tau^{a}\,\partial_{\mu}A^{h,a}_{\mu}
+\frac{b_{2}}{2}\,J\,A^{h,a}_{\mu}A^{h,a}_{\mu}+b_{3}\,\frac{\zeta}{2}\,J^{2}\nonumber\\
&&
+a_{1}\,A^{a}_{\mu}\frac{\delta\Sigma}{\delta A^{a}_{\mu}}
-ia_{1}\,b^{a}\,\partial_{\mu}A^{a}_{\mu}
-a_{1}\,\Omega^{a}_{\mu}\frac{\delta\Sigma}{\delta\Omega^{a}_{\mu}}
+a_{1}\,\bar{c}^{a}\,\partial_{\mu}\frac{\delta\Sigma}{\delta\Omega^{a}_{\mu}}
-a_{2}\,c^{a}\frac{\delta\Sigma}{\delta c^{a}}\nonumber\\
&&
+a_{2}\,L^{a}\frac{\delta\Sigma}{\delta L^{a}}
+f^{a}(\xi)\frac{\delta\Sigma}{\delta\xi^{a}}
-K^{a}\frac{\partial f^{a}}{\partial\xi^{b}}\frac{\delta\Sigma}{\delta K^{b}}\,\bigg)\,.
\label{count_rewritten}
\end{eqnarray}
where use has been made of  the explicit expressions of $\widehat{\mathcal{J}}^{a}_{\mu}$ 
and $\widehat{\Omega}^{a}_{\mu}$ given, respectively, in eqs.\eqref{J_hat},\eqref{omega_hat}. 
In order to analyze the  different terms of expression \eqref{count_rewritten}, we set
\begin{eqnarray}
\Sigma^{ct}&=&\sum_{n=1}^{7}\,\Sigma^{ct}_{n}\,,
\end{eqnarray}
with
\begin{eqnarray}
\Sigma^{ct}_{1}&=&\int d^{4}x\,\frac{a_{0}}{4g^{2}}\,F^{a}_{\mu\nu}F^{a}_{\mu\nu}\,,\nonumber\\
\Sigma^{ct}_{2}&=&\int d^{4}x\,b_{1}\,\mathcal{J}^{a}_{\mu}A^{h,a}_{\mu}\,,\nonumber\\
\Sigma^{ct}_{3}&=&\int d^{4}x\,b_{1}\,\tau^{a}\,\partial_{\mu}A^{h,a}_{\mu}\,,\nonumber\\
\Sigma^{ct}_{4}&=&\int d^{4}x\,\left(\frac{b_{2}}{2}\,J\,A^{h,a}_{\mu}A^{h,a}_{\mu}
+b_{3}\,\frac{\zeta}{2}\,J^{2}\right)\,,\nonumber\\
\Sigma^{ct}_{5}&=&\int d^{4}x\,(-ia_{1}\,b^{a}\,\partial_{\mu}A^{a}_{\mu})\,,\nonumber\\
\Sigma^{ct}_{6}&=&\int d^{4}x\,a_{1}\,\bar{c}^{a}\,\partial_{\mu}\frac{\delta\Sigma}{\delta\Omega^{a}_{\mu}}\,,\nonumber\\
\Sigma^{ct}_{7}&=&\int d^{4}x\,\left(a_{1}\,A^{a}_{\mu}\frac{\delta\Sigma}{\delta A^{a}_{\mu}}
-a_{1}\,\Omega^{a}_{\mu}\frac{\delta\Sigma}{\delta\Omega^{a}_{\mu}}
+a_{2}\,L^{a}\frac{\delta\Sigma}{\delta L^{a}}
+f^{a}(\xi)\frac{\delta\Sigma}{\delta\xi^{a}}
-K^{a}\frac{\partial f^{a}}{\partial\xi^{b}}\frac{\delta\Sigma}{\delta K^{b}}\right)\,.
\end{eqnarray}
By noticing that 
\begin{equation}
\frac{\partial\Sigma}{\partial g^{2}}=
\frac{\partial}{\partial g^{2}}\int d^{4}x\,\frac{1}{4g^{2}}\,F^{a}_{\mu\nu}F^{a}_{\mu\nu}=-\frac{1}{g^{2}}\int d^{4}x\,\frac{1}{4g^{2}}\,F^{a}_{\mu\nu}F^{a}_{\mu\nu}\,,
\end{equation}
the term $\Sigma^{ct}_{1}$ can be rewritten as 
\begin{equation}
\Sigma^{ct}_{1}=-a_{0}\,g^{2}\frac{\partial\Sigma}{\partial g^{2}}\,.
\label{count_1}
\end{equation}
Taking the variation of the action $\Sigma$ with respect to $\mathcal{J}^{a}_{\mu}$ and $\tau^{a}$,
\begin{equation}
\frac{\delta\Sigma}{\delta\mathcal{J}^{a}_{\mu}}=A^{h,a}_{\mu}\,,\qquad
\frac{\delta\Sigma}{\delta\tau^{a}}=\partial_{\mu}A^{h,a}_{\mu}\,,
\end{equation}
the terms $\Sigma^{ct}_{2}$ and $\Sigma^{ct}_{3}$ are rewritten as
\begin{eqnarray}
\Sigma_{2}^{count}&=&b_{1}\int d^{4}x\,\mathcal{J}^{a}_{\mu}\frac{\delta\Sigma}{\delta\mathcal{J}^{a}_{\mu}}\,,\nonumber\\
\Sigma_{3}^{count}&=&b_{1}\int d^{4}x\,\tau^{a}\frac{\delta\Sigma}{\delta\tau^{a}}\,.
\label{count_2_and_3}
\end{eqnarray}
Also, taking the variation of $\Sigma$ with respect to $J$,  we obtain
\begin{equation}
\frac{\delta\Sigma}{\delta J}=\frac{1}{2}\,A^{h,a}_{\mu}A^{h,a}_{\mu}+\zeta\,J\,,
\end{equation}
from which it follows that $\Sigma_{4}^{ct}$ takes the form 
\begin{equation}
\Sigma_{4}^{ct}=\int d^{4}x\,\left(b_{2}\,J\frac{\delta\Sigma}{\delta J}+(b_3-2b_2)\,\frac{\zeta}{2}\,J^{2}\right)\,.
\end{equation}
On the other hand, we also have that
\begin{equation}
\zeta\frac{\partial\Sigma}{\partial\zeta}=\int d^{4}x\,\frac{\zeta}{2}\,J^{2}\,.
\end{equation}
Thus,
\begin{equation}
\Sigma_{4}^{ct}=b_{2}\int d^{4}x\,J\frac{\delta\Sigma}{\delta J}+(b_3-2b_2)\,\zeta\frac{\partial\Sigma}{\partial\zeta}\,.
\label{count_4}
\end{equation}
Now, considering the gauge fixing equation \eqref{gauge_fixing}, we can rewrite $\Sigma_{5}^{ct}$ as
\begin{equation}
\Sigma^{ct}_{5}=\int d^{4}x\,\left(-a_{1}\,b^{a}\frac{\delta\Sigma}{\delta b^{a}}+a_1\,\alpha\,b^{a}b^{a}\right)\,.
\end{equation}
Furthermore, from 
\begin{equation}
2\alpha\frac{\partial\Sigma}{\partial\alpha}=\int d^{4}x\,\alpha\,b^{a}b^{a}\,,
\end{equation}
one gets
\begin{equation}
\Sigma^{ct}_{5}=-a_1\int d^{4}x\,b^{a}\frac{\delta\Sigma}{\delta b^{a}}
+2a_1\,\alpha\frac{\partial\Sigma}{\partial\alpha}\,.
\label{count_5}
\end{equation}
The term $\Sigma^{ct}_{6}$ can be immediately rewritten using the anti-ghost equation \eqref{aghost_eq} as:
\begin{equation}
\Sigma^{ct}_{6}=-a_1\int d^{4}x\, \bar{c}^{a}\frac{\delta\Sigma}{\delta\bar{c}^{a}}\,.
\label{count_6}
\end{equation}
Putting together all expressions, for the parametric form of the counterterm we obtain
\begin{eqnarray}
\Sigma^{ct}&=&-a_{0}\,g^{2}\frac{\partial\Sigma}{\partial g^{2}}
+(b_{3}-2b_{2})\,\zeta\frac{\partial\Sigma}{\partial\zeta}
+2a_{1}\,\alpha\frac{\partial\Sigma}{\partial\alpha}
+\int d^{4}x\,\bigg(a_{1}\,A^{a}_{\mu}\frac{\delta\Sigma}{\delta A^{a}_{\mu}}
-a_{1}\,b^{a}\frac{\delta\Sigma}{\delta b^{a}}-\nonumber\\
&&
-a_{1}\,\bar{c}^{a}\frac{\delta\Sigma}{\delta\bar{c}^{a}}
-a_{2}\,c^{a}\frac{\delta\Sigma}{\delta c^{a}}
+b_{1}\,\tau^{a}\frac{\delta\Sigma}{\delta\tau^{a}}
+f^{a}(\xi)\frac{\delta\Sigma}{\delta\xi^{a}}
-a_{1}\,\Omega^{a}_{\mu}\frac{\delta\Sigma}{\delta\Omega^{a}_{\mu}}
+a_{2}\,L^{a}\frac{\delta\Sigma}{\delta L^{a}}+\nonumber\\
&&
+b_{1}\,\mathcal{J}^{a}_{\mu}\frac{\delta\Sigma}{\delta\mathcal{J}^{a}_{\mu}}
+b_{2}\,J\frac{\delta\Sigma}{\delta J}
-K^{b}\frac{\partial f^{b}}{\partial\xi^{a}}\frac{\delta\Sigma}{\delta K^{a}}\bigg)\,,
\end{eqnarray}
which can be finally written as 
\begin{equation}
\Sigma^{ct}=\mathcal{R}\Sigma \,, \label{R}
\end{equation}
with $\mathcal{R}$ being the differential operator  
\begin{eqnarray}
\mathcal{R}&=&-a_{0}\,g^{2}\frac{\partial}{\partial g^{2}}
+(b_{3}-2b_{2})\,\zeta\frac{\partial}{\partial\zeta}
+2a_{1}\,\alpha\frac{\partial}{\partial\alpha}
+\int d^{4}x\,\bigg(a_{1}\,A^{a}_{\mu}\frac{\delta}{\delta A^{a}_{\mu}}
-a_{1}\,b^{a}\frac{\delta}{\delta b^{a}}\nonumber\\
&&
-a_{1}\,\bar{c}^{a}\frac{\delta}{\delta\bar{c}^{a}}
-a_{2}\,c^{a}\frac{\delta}{\delta c^{a}}
+b_{1}\,\tau^{a}\frac{\delta}{\delta\tau^{a}}
+f^{a}(\xi)\frac{\delta}{\delta\xi^{a}}
-a_{1}\,\Omega^{a}_{\mu}\frac{\delta}{\delta\Omega^{a}_{\mu}}
+a_{2}\,L^{a}\frac{\delta}{\delta L^{a}}\nonumber\\
&&
+b_{1}\,\mathcal{J}^{a}_{\mu}\frac{\delta}{\delta\mathcal{J}^{a}_{\mu}}
+b_{2}\,J\frac{\delta}{\delta J}
-K^{b}\frac{\partial f^{b}}{\partial\xi^{a}}\frac{\delta}{\delta K^{a}}\bigg)\,.   \label{R1} 
\end{eqnarray}
The usefulness of expression \eqref{R} relies on the fact that it immediately provides the redefinition of the fields, parameter and sources needed to show that the counterterm $\Sigma^{ct}$ can be in fact reabsorbed into the starting action, namely 
\begin{equation}
\Sigma(\Phi) + \varepsilon \Sigma^{ct}(\Phi) = \Sigma(\Phi_0) + O(\varepsilon^{2})\,, \label{stab}
\end{equation}
where $\varepsilon$ is an expansion parameter, $\Phi$ is a shorthand notation for the fields, parameters and sources, while $\Phi_0$ stands for the corresponding redefinitions.  From equation \eqref{R} it is apparent that the redefined fields, parameters and sources are given by 
\begin{equation}
\Phi_0=(1+\varepsilon\,\mathcal{R})\Phi\,.
\label{renorm}
\end{equation}
In fact, using \eqref{renorm}, it is almost immediate to prove that 
\begin{equation}
\Sigma[\Phi_0]=\Sigma[\Phi+\varepsilon\,\mathcal{R}\Phi]=\Sigma[\Phi]+\varepsilon\,\mathcal{R}(\Sigma)+O(\varepsilon^{2})\,,   \label{stab2}
\end{equation}
showing that the counterterm $\Sigma^{ct}$  can be reabsorbed into the starting action $\Sigma$.\\\\By direct inspection of equation \eqref{stab2}, for the renormalization factors one finds: 
\begin{eqnarray}
&A_{0}=Z^{1/2}_{A}\,A\,,\quad
b_{0}=Z^{1/2}_{b}\,b\,,\quad
c_{0}=Z^{1/2}_{c}\,c\,,\quad
\bar{c}_{0}=Z^{1/2}_{\bar{c}}\,\bar{c}\,,\quad
\xi^{a}_{0}=Z^{ab}_{\xi}(\xi)\,\xi^{b}\,,\quad
\tau_{0}=Z^{1/2}_{\tau}\,\tau\,,&\nonumber\\
&\Omega_{0}=Z_{\Omega}\,\Omega\,,\quad
L_{0}=Z_{L}\,L\,,\quad
K^{a}_0=Z_{K}^{ab}(\xi)\,K^{b}\,,\quad
J_{0}=Z_{J}\,J\,,\quad
\mathcal{J}_{0}=Z_{\mathcal{J}}\,\mathcal{J}\,,\quad&\nonumber\\
&g_{0}=Z_{g}\,g\,,\quad
\alpha_{0}=Z_{\alpha}\,\alpha\,,\quad
\zeta_{0}=Z_{\zeta}\,\zeta\,,&
\end{eqnarray}
where
\begin{eqnarray}
Z_{g}&=&1+\varepsilon\,a_{0}\,,\\
Z^{1/2}_{A}&=&1+\varepsilon\,a_{1}\,,\\
Z^{1/2}_{c}&=&1-\varepsilon\,a_{2}\,,\\
Z_{\mathcal{J}}&=&1+\varepsilon\,b_{1}\,,\\
Z_{J}&=&1+\varepsilon\,b_{2}\,,\\
Z_{\zeta}&=&1+\varepsilon\,(b_{3}-2b_{2})\,,\\
Z^{ab}_{\xi}&=&\delta^{ab}+\varepsilon\,f^{ab}_{3}\,,\label{Z_xi}\\
Z^{ab}_{K}&=&\delta^{ab}-\varepsilon\,\frac{\partial f^{b}}{\partial\xi^{a}}
\,\,=\,\,\delta^{ab}-\varepsilon\,\left(\frac{\partial f^{bc}_{3}}{\partial\xi^{a}}\,\xi^{c}+f^{ba}_{3}\right)  \;,
\label{Z_K}
\end{eqnarray}
and 
\begin{equation}
Z_{\alpha}=Z_{A}\,,\qquad
Z^{1/2}_{b}=Z^{1/2}_{\bar{c}}=Z_{\Omega}=Z^{-1/2}_{A}\,,\qquad
Z^{1/2}_{\tau}=Z_{\mathcal{J}}\,,\qquad
Z_{L}=Z^{-1/2}_{c}\,.
\end{equation}
Observe that in equations \eqref{Z_xi} and \eqref{Z_K} we have used the definition 
$f^{a}(\xi)=f^{ab}_{3}(\xi)\xi^{b}$ introduced in eq.\eqref{redef}. We also underline 
that, according to \eqref{Z_xi},\eqref{Z_K}, the renormalization factors of the 
Stueckelberg field $\xi^{a}$ and of the corresponding  source $K^{a}$ are nonlinear, 
{\it i.e.} they are power series in $\xi^a$. This is an expected feature,  due to the 
dimensionless character of the Stueckelberg field, a feature common to other 
renormalizable models displaying massles fields as, for example,  $N=1$ super Yang-Mills 
theory in superspace, see \cite{Piguet:1986ug}.

\section{The anomalous dimensions of $(A^h_\mu A^h_\mu)$ and of $A^{h}_{\mu}$} \label{andim}
Let us address now the issue of the anomalous dimensions of the operators   
$(A^{h,a}_{\mu}A^{h,a}_{\mu})$ and $A^{h,a}_{\mu}$. As a consequence of their gauge 
invariance, their anomalous dimensions turn out to be independent from the gauge 
parameter $\alpha$, a result which  can be established at the algebraic level through 
the use of the extended BRST technique \cite{Piguet:1995er}. (See also the recent proof 
given in \cite{Capri:2016aqq}.) 
\\\\In particular, due to its $\alpha$-independence, the anomalous dimension of 
$(A^{h,a}_{\mu}A^{h,a}_{\mu})$ is the same as that computed in the Landau gauge, 
{\it i.e.} for $\alpha=0$. Moreover, taking into account that, in the Landau gauge, 
the operator $(A^{h,a}_{\mu}A^{h,a}_{\mu})$ reduces to $(A^{a}_{\mu}A^{a}_{\mu})$, we 
expect that the anomalous dimension $\gamma_{(A^{h})^2}$ of $(A^{h,a}_{\mu}A^{h,a}_{\mu})$
should be equal to the anomalous dimension $\gamma_{A^{2}}\Big|_{\rm Landau}$ of the 
operator $(A^{a}_{\mu}A^{a}_{\mu})$  in the Landau gauge, namely 
\begin{equation}
\gamma_{(A^{h})^2}= \gamma_{A^{2}}\Big|_{\rm Landau}  =-\left(\frac{\beta(a)}{a}+\gamma^{\mathrm{Landau}}_{A}(a)\right)\,,\qquad
a=\frac{g^{2}}{16\pi^{2}}\,,
\label{gamma_AA}
\end{equation}
where $(\beta(a), \gamma^{\rm Landau}_A(a))$ denote, respectively, the $\beta$-function and 
the anomalous dimension of the gauge field $A_\mu$ in the Landau gauge\footnote{For an all order algebraic proof of the relationship 
\begin{equation}
\gamma_{A^{2}}\Big|_{\rm Landau}  =-\left(\frac{\beta(a)}{a}+\gamma^{\mathrm{Landau}}_{A}(a)\right)\,,\qquad
a=\frac{g^{2}}{16\pi^{2}}\,,\nonumber
\label{gamma_AA_L}
\end{equation}
see  \cite{Dudal:2002pq}.}.  \\\\A similar property is expected in the case of the operator $A^{h}_{\mu}$, namely 
\begin{equation}
\gamma_{A^{h}}=\gamma_{A^{h}}\Big|_{\alpha=0}=\gamma^{\mathrm{Landau}}_{A}(a)\,,   
\label{gamma_A}
\end{equation}
{\it i.e.} the anomalous dimension of $A^{h}_{\mu}$ should equal  that  of the gauge field $A^{a}_{\mu}$ in the Landau gauge. Therefore, both $\gamma_{(A^{h})^2}$ and $\gamma_{A^{h}}$ would not be independent parameters of the theory. \\\\Let us give a formal proof of equations \eqref{gamma_AA} and \eqref{gamma_A} by making use of the renormalization group equation (RGE) which, owing to the renormalizability and to the BRST invariance of the theory, reads 
\begin{eqnarray}
& & \mu\frac{\partial\Gamma}{\partial\mu} +  \beta_{g^{2}}\,\frac{\partial\Gamma}{\partial g^{2}}
-\gamma_{A}\,\mathcal{N}_{A}\Gamma
-\gamma_{c}\,\mathcal{N}_{c}\Gamma
-\gamma_{(A^{h})^2}\int d^{4}x\, J\frac{\delta\Gamma}{\delta J}\nonumber\\
&&
-\gamma_{A^{h}}\int d^{4}x\,\left(\mathcal{J}^{a}_{\mu}\frac{\delta\Gamma}{\delta\mathcal{J}^{a}_{\mu}}+\tau^{a}\frac{\delta\Gamma}{\delta\tau^{a}}\right)
-\int d^{4}x\left(\gamma_{\xi}^{ab}(\xi)\xi^{b}\frac{\delta\Gamma}{\delta\xi^{a}}+
\gamma_{K}^{ab}(\xi)K^{b}\frac{\delta\Gamma}{\delta K^{a}}\right) = 0 \,,
\label{RGE}
\end{eqnarray}
where
\begin{eqnarray}
\mathcal{N}_{A}&=&\int d^{4}x\left(A^{a}_{\mu}\frac{\delta}{\delta A^{a}_{\mu}}
-b^{a}\frac{\delta}{\delta b^{a}}
-\bar{c}^{a}\frac{\delta}{\delta \bar{c}^{a}}
-\Omega^{a}_{\mu}\frac{\delta}{\delta \Omega^{a}_{\mu}}\right)
+2\alpha\,\frac{\partial}{\partial\alpha}\,,\nonumber\\
\mathcal{N}_{c}&=&\int d^{4}x\,\left(
c^{a}\frac{\delta}{\delta c^{a}}
-L^{a}\frac{\delta}{\delta L^{a}}\right)\,,\nonumber\\
\gamma^{ab}_{\xi}&=&(Z^{-1}_{\xi})^{ac}\,\mu\frac{\partial}{\partial\mu}Z^{cb}_{\xi}\,,\nonumber\\
\gamma^{ab}_{K}&=&(Z^{-1}_{K})^{ac}\,\mu\frac{\partial}{\partial\mu}Z^{cb}_{K}\,.
\end{eqnarray}
Let us act now on the RGE with the test operator 
\begin{equation}
\frac{\delta^{2}}{\delta \mathcal{J}^{a}_{\mu}(x)\delta \mathcal{J}^{b}_{\nu}(y)} \,,
\end{equation}
and set all fields and sources equal to zero. A simple algebraic calculation gives 
\begin{eqnarray}
0&=&\mu\,\frac{\partial}{\partial{\mu}}\langle A^{h,a}_{\mu}(x)A^{h,b}_{\nu}(y)\rangle
+\beta_{g^{2}}\,\frac{\partial}{\partial g^{2}}\langle A^{h,a}_{\mu}(x)A^{h,b}_{\nu}(y)\rangle\nonumber\\
&&-2\gamma_{A}\,\alpha\frac{\partial}{\partial \alpha}\langle A^{h,a}_{\mu}(x)A^{h,b}_{\nu}(y)\rangle
-2\gamma_{A^{h}}\,\langle A^{h,a}_{\mu}(x)A^{h,b}_{\nu}(y)\rangle\,.
\end{eqnarray}
Moreover, due to the $\alpha$-independence of the gauge-invariant correlation function $\langle A^{h,a}_{\mu}(x)A^{h,b}_{\nu}(y)\rangle$, it follows that
\begin{equation}
\frac{\partial}{\partial \alpha}\langle A^{h,a}_{\mu}(x)A^{h,b}_{\nu}(y)\rangle=0\,.   \label{alphaind}
\end{equation}
Thus,
\begin{equation}
\mu\,\frac{\partial}{\partial{\mu}}\langle A^{h,a}_{\mu}(x)A^{h,b}_{\nu}(y)\rangle
+\beta_{g^{2}}\,\frac{\partial}{\partial g^{2}}\langle A^{h,a}_{\mu}(x)A^{h,b}_{\nu}(y)\rangle
-2\gamma_{A^{h}}\,\langle A^{h,a}_{\mu}(x)A^{h,b}_{\nu}(y)\rangle=0\,.
\end{equation}
In addition, from \eqref{alphaind} we can make direct use of the Landau gauge, namely 
\begin{equation}
\langle A^{h,a}_{\mu}(x)A^{h,b}_{\nu}(y)\rangle
=\langle A^{h,a}_{\mu}(x)A^{h,b}_{\nu}(y)\rangle_{\alpha=0}
=\langle A^{h,a}_{\mu}(x)A^{h,b}_{\nu}(y)\rangle_{\mathrm{Landau}}\,.
\end{equation}
Therefore
\begin{equation}
\langle A^{h,a}_{\mu}(x)A^{h,b}_{\nu}(y)\rangle=\frac{\int [D\phi]\,A^{h,a}_{\mu}(x)A^{h,b}_{\nu}(y)\,e^{-S_{(\alpha, m^2=0)}}}{\int [D\phi]\,e^{-S_{(\alpha, m^2=0)}}}\,,
\end{equation}
with $[D\phi]\equiv DADbDcD\bar{c}D\xi D\tau$ and
\begin{equation}
S_{(\alpha,m^{2}=0)}=\int d^{4}x\,\left(\frac{1}{4g^{2}}\,F^{a}_{\mu\nu}F^{a}_{\mu\nu}
+ib^{a}\,\partial_{\mu}A^{a}_{\mu}
+\bar{c}^{a}\,\partial_{\mu}D^{ab}_{\mu}c^{b}+\tau^{a}\,\partial_{\mu}A^{h,a}_{\mu}\right)\,.
\end{equation}
Integrating out the fields $(\tau,b,c,\bar{c})$, we get 
\begin{equation}
\langle A^{h,a}_{\mu}(x)A^{h,b}_{\nu}(y)\rangle=\frac{\int DAD\xi\,\delta(\partial_{\mu}A^{h}_{\mu})\delta(\partial_{\mu}A_{\mu})\det(-\partial\cdot D)\,A^{h,a}_{\mu}(x)A^{h,b}_{\nu}(y)\,e^{-S_{\mathrm{YM}}}}{\int DAD\xi\,\delta(\partial_{\mu}A^{h}_{\mu})\delta(\partial_{\mu}A_{\mu})\det(-\partial\cdot D)\,e^{-S_{\mathrm{YM}}}}\,.
\end{equation}
Employing the result given in Appendix \eqref{apb}, see eqs.\eqref{hh2},\eqref{phi0}, the equation $\partial_\mu A^h_\mu=0$ can be solved iteratively for $\xi^a$ yielding 
\begin{equation}
\xi =\frac{1}{\partial ^{2}}\partial _{\mu }A_{\mu }+i\frac{g}{\partial ^{2}%
}\left[ \partial A,\frac{\partial A}{\partial ^{2}}\right] +i\frac{g}{%
\partial ^{2}}\left[ A_{\mu },\partial _{\mu }\frac{\partial A}{\partial ^{2}%
}\right] +\frac{i}{2}\frac{g}{\partial ^{2}}\left[ \frac{\partial A}{%
\partial ^{2}},\partial A\right] +O(A^{3})\;,  \label{xi1}
\end{equation}
so that we can integrate over $\xi^{a}$, obtaining 
\begin{equation}
\langle A^{h,a}_{\mu}(x)A^{h,b}_{\nu}(y)\rangle=\frac{\int DA\,\delta(\partial_{\mu}A_{\mu})\det(-\partial\cdot D)\,A^{h,a}_{\mu}(x)A^{h,b}_{\nu}(y)\,e^{-S_{\mathrm{YM}}}}{\int DA\,\delta(\partial_{\mu}A_{\mu})\det(-\partial\cdot D)\,e^{-S_{\mathrm{YM}}}}\,, \label{qa1}
\end{equation}
where $A^h_\mu$ is now given by, see eq.\eqref{minn2} of Appendix \eqref{apb}, 
\begin{eqnarray}
A_{\mu }^{h} &=&A_{\mu }-\frac{1}{\partial ^{2}}\partial _{\mu }\partial A-ig%
\frac{\partial _{\mu }}{\partial ^{2}}\left[ A_{\nu },\partial _{\nu }\frac{%
\partial A}{\partial ^{2}}\right] -i\frac{g}{2}\frac{\partial _{\mu }}{%
\partial ^{2}}\left[ \partial A,\frac{1}{\partial ^{2}}\partial A\right]
\nonumber \\
&+&ig\left[ A_{\mu },\frac{1}{\partial ^{2}}\partial A\right] +i\frac{g}{2}%
\left[ \frac{1}{\partial ^{2}}\partial A,\frac{\partial _{\mu }}{\partial
^{2}}\partial A\right] +O(A^{3})\;.  \label{qa2}
\end{eqnarray}
However, due to the presence in  eq.\eqref{qa1} of the delta function $\delta(\partial_{\mu}A_{\mu})$, all terms containing a divergence $\partial A$ vanish, namely 
\begin{eqnarray}
\langle A^{h,a}_{\mu}(x)A^{h,b}_{\nu}(y)\rangle&=&\frac{\int DA\,\delta(\partial_{\mu}A_{\mu})\det(-\partial\cdot D)\,A^{h,a}_{\mu}(x)A^{h,b}_{\nu}(y)\,e^{-S_{\mathrm{YM}}}}{\int DA\,\delta(\partial_{\mu}A_{\mu})\det(-\partial\cdot D)\,e^{-S_{\mathrm{YM}}}}\nonumber\\
&=&\frac{\int DA\,\delta(\partial_{\mu}A_{\mu})\det(-\partial\cdot D)\,A^{a}_{\mu}(x)A^{b}_{\nu}(y)\,e^{-S_{\mathrm{YM}}}}{\int DA\,\delta(\partial_{\mu}A_{\mu})\det(-\partial\cdot D)\,e^{-S_{\mathrm{YM}}}}\nonumber\\
&=&\langle A^{a}_{\mu}(x)A^{b}_{\nu}(y)\rangle_{\mathrm{Landau}}\,.
\end{eqnarray}
Thus, the RGE for the correlation function $\langle A^{h,a}_{\mu}(x)A^{h,b}_{\nu}(y)\rangle$ becomes
\begin{equation}
\mu\,\frac{\partial}{\partial{\mu}}\langle A^{a}_{\mu}(x)A^{b}_{\nu}(y)\rangle_{\mathrm{Landau}}
+\beta_{g^{2}}\,\frac{\partial}{\partial g^{2}}\langle A^{a}_{\mu}(x)A^{b}_{\nu}(y)\rangle_{\mathrm{Landau}}
-2\gamma_{A^{h}}\,\langle A^{a}_{\mu}(x)A^{b}_{\nu}(y)\rangle_{\mathrm{Landau}}=0\,, 
\end{equation}
which proves equation \eqref{gamma_A}. Of course, the same reasoning can be applied to  equation \eqref{gamma_AA}.

\section{Conclusions}

In this work we have provided a study of the gauge-invariant non-local operator $A_{\min }^{2}$ 
\begin{equation}
A_{\min }^{2} =\mathrm{Tr}\int d^{4}x\,A_{\mu }^{h}A_{\mu }^{h}\;,   \label{cc1}
\end{equation}
with $A_\mu^h$ the transverse configuration, $\partial_\mu A^h_\mu=0$, given in expression \eqref{min0}. \\\\Despite the highly non-local character, we have shown that a fully local set up for both operators 
$(A^h_\mu A^h_\mu)$ and $A^h_\mu$ can be constructed, giving rise to a local and BRST-invariant action $S$, eq.\eqref{S_local}. The main tool in order to achieve such a local formulation has been the introduction of an auxiliary Stueckelberg field $\xi^a$, eqs.\eqref{local_Ah},\eqref{hxi}.  \\\\As pointed out in Section \eqref{local_framework}, the transversality condition,  $\partial_\mu A^h_\mu=0$, plays an important role, giving rise to deep differences between our formulation and the conventional Stueckelberg one, which is known to be non-renormalizable. Unlike the conventional Stueckelberg formulation, the novel action $S$,  eq.\eqref{S_local}, has been proven to be renormalizable to all orders, as shown in details in Sect.\eqref{Renormalizability}. Furthermore, owing to the gauge invariance of $(A^h_\mu A^h_\mu)$ and $A^h_\mu$, the corresponding anomalous dimensions, $(\gamma_{(A^{h})^2}, \gamma_{A^{h}})$, turn out to be independent from the gauge parameter $\alpha$ entering the gauge fixing condition, 
 being given by 
\begin{eqnarray}
\gamma_{(A^{h})^2} &= &  \gamma_{A^{2}}\Big|_{\rm Landau}  =-\left(\frac{\beta(a)}{a}+\gamma^{\mathrm{Landau}}_{A}(a)\right)\,,\qquad
a=\frac{g^{2}}{16\pi^{2}}\,,   \nonumber \\
\gamma_{A^{h}}& = & \gamma_{A^{h}}\Big|_{\alpha=0}=\gamma^{\mathrm{Landau}}_{A}(a)\,,   \label{ad} 
\end{eqnarray} 
where $(\beta(a), \gamma^{\rm Landau}_A(a))$ denote, respectively, the $\beta$-function and 
the anomalous dimension of the gauge field $A_\mu$ in the Landau gauge. We see therefore that $(\gamma_{(A^{h})^2}, \gamma_{A^{h}})$ are not independent parameters of the theory. \\\\The present results can open the road to several  future investigations. For instance, the possibility of having at our disposal a local and renormalizable framework might enable us to investigate the formation, through the computation of the effective potential \cite
{Verschelde:2001ia,Dudal:2003vv,Browne:2003uv,Browne:2004mk,Gracey:2004bk}, of the gauge-invariant dimension-two condensate $\langle A^h_\mu A^h_\mu \rangle$. This result might yield a better understanding, within a manifestly BRST-invariant set up,  of the relevance of the  condensate $\langle A^h_\mu A^h_\mu \rangle$ for the formation of the dynamical gluon mass  \cite
{Verschelde:2001ia,Dudal:2003vv,Browne:2003uv,Browne:2004mk,Gracey:2004bk} as well for the analysis of the  $\frac{1}{Q^2}$ corrections in the gluon correlation functions within the OPE expansion, as reported in \cite{Boucaud:2001st,Boucaud:2002nc,Boucaud:2005rm,RuizArriola:2004en,Furui:2005bu,Boucaud:2005xn,Boucaud:2008gn,Pene:2011kg,Boucaud:2010gr,Blossier:2010ky,Boucaud:2011eh,Blossier:2011tf,Blossier:2013te} in the case of the Landau gauge. \\\\Another topic worth to be mentioned is the study of the BRST-invariant and $\alpha$-independent correlation function 
\begin{equation}
\langle A^h_\mu(x) A^h_\mu(y) \rangle  \;, \label{ahah}
\end{equation}
within the local present set up. Due to its $\alpha$-independence, expression \eqref{ahah} can be seen as the natural generalization, in the case of the covariant linear gauges, of the two-point function  $\langle A_\mu(x) A_\mu(y) \rangle_{\rm Landau}$ studied in the renormalizable massive Yang-Mills model in the Landau gauge considered in  \cite{Tissier:2010ts,Tissier:2011ey}. As such, expression \eqref{ahah} might provide information about the occurrence of positivity violation, already observed in the Landau gauge \cite{Tissier:2010ts,Tissier:2011ey}.  In this sense, expression  \eqref{ahah} might be regarded as a powerful and practical tool to detect the positivity violation, in linear covariant gauges, of the two-point gluon correlation function within a BRST-invariant formulation.


\section*{Acknowledgements.}
The Conselho Nacional de Desenvolvimento Cient\'{i}fico e
Tecnol\'{o}gico (CNPq-Brazil), the Faperj, Funda{\c{c}}{\~{a}}o de
Amparo {\`{a}} Pesquisa do Estado do Rio de Janeiro, the SR2-UERJ
and the Coordena{\c{c}}{\~{a}}o de Aperfei{\c{c}}oamento de
Pessoal de N{\'\i}vel Superior (CAPES) are gratefully acknowledged
for financial support.

\appendix

\section{A review on the renormalization of the Yang-Mills action in linear covariant gauges}
\label{review}
When the mass parameter $m^2$ is set to zero, the action $S$ in eq.\eqref{nact} reduces to 

\begin{equation}
S_{m^2=0} = S_{FP} + \int d^4x \left(\tau^{a}\,\partial_{\mu}A^{h,a}_{\mu} \right) \; 
  \label{nact1} 
\end{equation}
where $S_{FP}$ is 
  \begin{equation}
S_{FP} =  \int d^{4}x\,\bigg(\frac{1}{4g^2}\,F^{a}_{\mu\nu}F^{a}_{\mu\nu}+\frac{\alpha}{2}\,b^{a}b^{a}
+ib^{a}\,\partial_{\mu}A^{a}_{\mu}
+\bar{c}^{a}\,\partial_{\mu}D^{ab}_{\mu}c^{b}  \;, 
\bigg)
  \label{fpact} 
\end{equation} 
{\it i.e.}  $S_{m^2=0}$ coincides, modulo the term  $\int d^4x\;  \tau^{a}\,\partial_{\mu}A^{h,a}_{\mu} $, with the usual Faddeev-Popov action of the linear covariant gauge. Evidently, the action $S_{m^2=0}$ is left invariant by the BRST transformations given in eqs.\eqref{nbrst},\eqref{nxibrst}
\begin{equation}
s S_{m^2=0} =0 \;. \label{inact1}
\end{equation}
Nevertheless, when ${m^2=0}$, the additional term $\int d^4x\;  \tau^{a}\,\partial_{\mu}A^{h,a}_{\mu} $ has no consequences on the evaluation of the Green functions of the elementary fields $(A_\mu, b, c, \bar c)$. More precisely, it turns out that the correlation functions 
$\langle A_{\mu_1}(x_1) ...... A_{\mu_n}(x_n) \rangle_{S_{m^2=0}} $ evaluated with the action $S_{m^2=0}$ coincide with those computed with the Faddeev-Popov action 
$S_{FP}$, namely 
\begin{equation}
\langle A_{\mu_1}(x_1) ...... A_{\mu_n}(x_n) \rangle_{S_{m^2=0}}  = \langle A_{\mu_1}(x_1) ...... A_{\mu_n}(x_n) \rangle_{S_{FP}}   \;.  \label{st}
\end{equation}
The statement \eqref{st} can be  checked by means of the functional integral. Let us consider expression 
\begin{equation}
\langle A_{\mu_1}(x_1) ...... A_{\mu_n}(x_n) \rangle_{S_{m^2=0}}  = \frac{ \int [D\phi]  \; A_{\mu_1}(x_1) ...... A_{\mu_n}(x_n)\; e^{-S_{m^2=0} }}{\int [D\phi]  \; e^{-S_{m^2=0}} }     \;, \label{st1}
\end{equation}
where $[D\phi] $ stands for integration over all fields, {\it i.e.} $[D\phi]= DA Db Dc D{\bar c} D\xi D\tau$. Integrating over the field $\tau$, one gets 
\begin{equation}
\langle A_{\mu_1}(x_1) ...... A_{\mu_n}(x_n) \rangle_{S_{m^2=0}}  = \frac{ \int DADbDcD\bar{c}D\xi \; \delta(\partial_\mu A^h_\mu)   \; A_{\mu_1}(x_1) ...... A_{\mu_n}(x_n)\; e^{-S_{FP }}}{\int DADbDcD\bar{c}D\xi \; \delta(\partial_\mu A^h_\mu) \; e^{-S_{FP}} }     \;, \label{st2}
\end{equation}
Making use of the result given in Appendix \eqref{apb}, see eqs.\eqref{hh2},\eqref{phi0}, the equation $\partial_\mu A^h_\mu=0$ can be solved iteratively for $\xi^a$ yielding 
\begin{equation}
\xi =\frac{1}{\partial ^{2}}\partial _{\mu }A_{\mu }+i\frac{g}{\partial ^{2}%
}\left[ \partial A,\frac{\partial A}{\partial ^{2}}\right] +i\frac{g}{%
\partial ^{2}}\left[ A_{\mu },\partial _{\mu }\frac{\partial A}{\partial ^{2}%
}\right] +\frac{i}{2}\frac{g}{\partial ^{2}}\left[ \frac{\partial A}{%
\partial ^{2}},\partial A\right] +O(A^{3})\;,  \label{xi1}
\end{equation}
so that expression \eqref{st2} can be written as 
\begin{equation}
\langle A_{\mu_1}(x_1) ...... A_{\mu_n}(x_n) \rangle_{S_{m^2=0}}  = \frac{ \int DADbDcD\bar{c}D\xi \;  \delta(\xi-[\mbox{power series in $A$}])  \; A_{\mu_1}(x_1) ...... A_{\mu_n}(x_n)\; e^{-S_{FP }}}{\int DADbDcD\bar{c}D\xi \; \delta(\xi-[\mbox{power series in $A$}]) \; e^{-S_{FP}} }     \;.  \label{st22}
\end{equation}
Observing now that the Faddeev-Popov action $S_{FP}$, eq.\eqref{nact1}, does not contain any dependence from the Stueckelberg field, it follows that the integration over the variable $\xi$ in equation \eqref{st2} is straightforward, giving 
\begin{equation}
\langle A_{\mu_1}(x_1) ...... A_{\mu_n}(x_n) \rangle_{S_{m^2=0}}  = \frac{ \int DADbDcD\bar{c}   \; A_{\mu_1}(x_1) ...... A_{\mu_n}(x_n)\; e^{-S_{FP }}}{\int DADbDcD\bar{c} \; e^{-S_{FP}} }     \;,  \label{st3}
\end{equation}
which proves the statement \eqref{st}. The same reasoning applies as well to other Green's functions containing the elementary fields $(b,c, {\bar c})$. In summary, all Green's function of the elementary fields $(A_\mu, b, c, {\bar c})$ evaluated with the  action \eqref{nact1}  are exactly the same as those computed with the Faddeev-Popov action  \eqref{fpact}. \\\\In particular, from this result it follows that the action  \eqref{nact1} is renormalizable, the most general counterterm being given, modulo terms in the variable $\tau$, by the usual counterterm of the linear covariant gauges.  \\\\Let us give a closer look at the possible local BRST-invariant counterterm $S^{ct}_{m^2=0}$ affecting the action  $S_{m^2=0}$ at the quantum level. $S^{ct}_{m^2=0}$  is a local integrated quantity in the fields bounded by dimension four. Moreover, it is useful to notice that, besides the BRST invariance, eq.\eqref{inact1}, the action $S_{m^2=0}$ is constrained by the additional Ward identity 
\begin{equation}
\int d^4x \frac{\delta S_{m^2=0}}{\delta \tau^a} = 0 \;, \label{wiact1}
\end{equation}
which implies that the variable $\tau$ can enter only through a space-time derivative, {\it i.e.} $\partial_\mu \tau^a$. Therefore, owing to the previous considerations, and taking into account that the field $\tau$ has dimension two, for the counterterm $S^{ct}_{m^2=0}$ we write 
\begin{equation}
S^{ct}_{m^2=0} = S^{ct}_{FP} - \int d^4x\; (\partial_\mu \tau^a)  \mathcal{O}^{a}_{\mu}(A,\xi)   \;, \label{cc1}
\end{equation}  
where $S^{ct}_{FP}$ is the usual local BRST-invariant counterterm of the Faddeev-Popov action in linear covariant gauges and where $\mathcal{O}^{a}_{\mu}(A,\xi) $ is a local quantity of dimension 1. From BRST invariance, we immediately get 
\begin{equation}
s \mathcal{O}^{a}_{\mu}(A,\xi) = 0 \;, 
\end{equation}
whose general solution, see eqs.\eqref{o1}-\eqref{Omu}, is 
\begin{equation}
\mathcal{O}^{a}_{\mu}(A,\xi) = b_1 \,A^{h,a}_{\mu}\,,     \label{Omuact1}
\end{equation}
with $b_1$ being an arbitrary coefficient. Thus, for the most general counterterm corresponding to $S|_{m^2=0}$ we have 
\begin{equation}
S^{ct}_{m^2=0} = S^{ct}_{FP} - b_1 \int d^4x\; (\partial_\mu \tau^a) A^{h,a}_{\mu}  \;. \label{countact1}
\end{equation}
Let us end this subsection by providing the expression  of the Faddeev-Popov counterterm  $S^{ct}_{FP}$, as derived form the algebraic renormalization procedure  \cite{Piguet:1995er}. 

\subsection{Renormalizability of the Faddeev-Popov  action in linear covariant gauges}

Following \cite{Piguet:1995er}, in order to determine the most general invariant counterterm $S^{ct}_{FP}$ affecting the Faddeev-Popov action in linear covariant gauges, eq.\eqref{fpact}, we start from the complete classical action 
\begin{equation}
\Sigma_{0}=S_{FP}+\int d^{4}x\,\left(-\Omega^{a}_{\mu}\,D^{ab}_{\mu}c^{b}+\frac{1}{2}f^{abc}L^{a}c^{b}c^{c}\right)\,,
\end{equation}
where we have introduced the external sources $(\Omega^{a}_\mu, L^a)$ coupled to the non-linear BRST variations of the fields $(A^a_\mu,c^a)$, see eqs.\eqref{nbrst},\eqref{nxibrst}. \\\\The action $\Sigma_0$ obeys the following set of Ward identities \cite{Piguet:1995er}:
\begin{eqnarray}
&\displaystyle\int d^{4}x\,\left(
\frac{\delta\Sigma_{0}}{\delta\Omega^{a}_{\mu}}
\frac{\delta\Sigma_{0}}{\delta A^{a}_{\mu}}
+\frac{\delta\Sigma_{0}}{\delta L^{a}}
\frac{\delta\Sigma_{0}}{\delta c^{a}}
+ib^{a}\,\frac{\delta\Sigma_{0}}{\delta\bar{c}^{a}}
\right)=0\,,&\\
&\displaystyle\frac{\delta\Sigma_{0}}{\delta b^{a}}=i\partial_{\mu}A^{a}_{\mu}+\alpha b^{a}\,,&\\
&\displaystyle\frac{\delta\Sigma_{0}}{\delta\bar{c}^{a}}+\partial_{\mu}\frac{\delta\Sigma_{0}}{\delta\Omega^{a}_{\mu}}=0\,,& 
\end{eqnarray} 
from which it turns out  \cite{Piguet:1995er} that the most general local invariant counterterm $\Sigma^{ct}_{0}$ contains three free parameters $(a_0, a_1, a_2)$, being given by the expression: 
\begin{eqnarray}
\Sigma^{ct}_{0}=a_0\int d^{4}x\,\frac{1}{4g^{2}}\,F^{a}_{\mu\nu}F^{a}_{\mu\nu}
+\mathcal{B}_{\Sigma_0}\int d^{4}x\,\left( a_{1}\,(\Omega^{a}_{\mu}+\partial_{\mu}\bar{c}^{a})A^{a}_{\mu}
+a_{2}\,L^{a}c^{a}\right)\,,\label{count_zero}
\end{eqnarray}
where $\mathcal{B}_{\Sigma_0}$ is the nilpotent linearized Slavnov-Taylor operator
\begin{equation}
\mathcal{B}_{\Sigma_0}=\int d^{4}x\,\left(
\frac{\delta\Sigma_{0}}{\delta\Omega^{a}_{\mu}}
\frac{\delta}{\delta A^{a}_{\mu}}
+\frac{\delta\Sigma_{0}}{\delta A^{a}_{\mu}}
\frac{\delta}{\delta\Omega^{a}_{\mu}}
+\frac{\delta\Sigma_{0}}{\delta L^{a}}
\frac{\delta}{\delta c^{a}}
+\frac{\delta\Sigma_{0}}{\delta c^{a}}
\frac{\delta}{\delta L^{a}}
+ib^{a}\frac{\delta}{\delta\bar{c}^{a}}
\right)\,, 
\end{equation}
\begin{equation}
\mathcal{B}_{\Sigma_0} \mathcal{B}_{\Sigma_0} =0    \;. \label{lnop}
\end{equation} 
Expression \eqref{count_zero} can be conveniently written in parametric form  \cite{Piguet:1995er} as 
\begin{eqnarray}
\Sigma^{ct}_{0}&=&-a_{0}\,g^{2}\frac{\partial\Sigma_{0}}{\partial g^{2}}
+2\alpha\,a_1\,\frac{\partial\Sigma_0}{\partial\alpha}+\int d^{4}x\,\bigg(
a_1\,A^{a}_{\mu}\frac{\delta\Sigma_{0}}{\delta A^{a}_{\mu}}
-a_1\,b^{a}\frac{\delta\Sigma_{0}}{\delta b^{a}}
-a_1\,\bar{c}^{a}\frac{\delta\Sigma_{0}}{\delta \bar{c}^{a}}\nonumber\\
&&
-a_1\,\Omega^{a}_{\mu}\frac{\delta\Sigma_{0}}{\delta \Omega^{a}_{\mu}}
-a_2\,c^{a}\frac{\delta\Sigma_{0}}{\delta c^{a}}
+a_2\,L^{a}\frac{\delta\Sigma_{0}}{\delta L^{a}}\bigg)\,,  \label{pffp}
\end{eqnarray}
which is suitable for establishing the renormalizablity of the starting action $\Sigma_{0}$, {\it i.e.} to check that $\Sigma^{ct}_{0}$ can be reabsorbed in $\Sigma_{0}$ through a redefinition of the fields, parameters and sources, according to 
\begin{equation}
\Sigma_{0}[A,b,c,\bar{c},\Omega,L,g^{2},\alpha]+\varepsilon\,\Sigma^{ct}_{0}
=\Sigma_{0}[A_0,b_0,c_0,\bar{c}_0,\Omega_0,L_0,g^{2}_0,\alpha_0]+O(\varepsilon^{2})\,,   \label{stab}
\end{equation}
with $\varepsilon$ stands for an expansion parameter and where  the label ``0'' denotes the redefined parameters, fields and sources. By direct inspection of equation \eqref{stab}, it follows that the counterterm $\Sigma^{ct}_{0}$ can be reabsorbed through the following redefinitions: 
\begin{equation}
g^{2}_0=Z_{g^{2}}\,g^{2}\,,\qquad
A_0=Z^{1/2}_{A}\,A\,,\qquad
c_0=Z^{1/2}_{c}\,c\,,
\end{equation}
with
\begin{eqnarray}
Z_{g^{2}}&=&1-\varepsilon\,a_0\,,\nonumber\\
Z^{1/2}_A &=&1+\varepsilon\,a_1\,,\nonumber\\
Z^{1/2}_{c}&=&1-\varepsilon\,a_2
\end{eqnarray}
and
\begin{eqnarray}
\alpha_0&=&Z_A\,\alpha\,,\nonumber\\
b_0&=&Z^{-1/2}_{A}\,b\,,\nonumber\\
\bar{c}_{0}&=&Z^{-1/2}_{A}\,\bar{c}\,,\nonumber\\
\Omega_{0}&=&Z^{-1/2}_{A}\,\Omega\,,\nonumber\\
L_0&=&Z^{-1/2}_{c}\,L\,,
\end{eqnarray} 
exhibiting the multiplicative all orders renormalizability of the Faddeev-Popov action in linear covariant gauges. \\\\Finally, setting the external sources $(\Omega^a_\mu, L^a)$ to zero, for the counterterm $S^{ct}_{FP}$, eq.\eqref{countact1}, one gets 
\begin{eqnarray}
S^{ct}_{FP} &=&  \Sigma^{ct}_{0}|_{\Omega=L=0}  \\  \nonumber 
&=&  -a_{0}\,g^{2}\frac{\partial S_{FP}}{\partial g^{2}}
+2\alpha\,a_1\,\frac{\partial S_{FP}}{\partial\alpha}+\int d^{4}x\,\bigg(
a_1\,A^{a}_{\mu}\frac{\delta S_{FP}}{\delta A^{a}_{\mu}}
-a_1\,b^{a}\frac{\delta S_{FP}}{\delta b^{a}}
-a_1\,\bar{c}^{a}\frac{\delta S_{FP}}{\delta \bar{c}^{a}}  -a_2\,c^{a}\frac{\delta S_{FP}}{\delta c^{a}}
\bigg)\,. \label{fpct}
\end{eqnarray}

\section{Properties of the functional $f_{A}[u]$.} \label{apb}In this
 Appendix we recall some useful properties of the functional
$f_{A}[u]$
\begin{equation}
f_{A}[u]\equiv \mathrm{Tr}\int d^{4}x\,A_{\mu }^{u}A_{\mu
}^{u}=\mathrm{Tr}\int d^{4}x\left( u^{\dagger }A_{\mu
}u+\frac{i}{g}u^{\dagger }\partial _{\mu }u\right) \left(
u^{\dagger }A_{\mu }u+\frac{i}{g}u^{\dagger }\partial _{\mu
}u\right) \;. \label{fa}
\end{equation}
For a given gauge field configuration $A_{\mu }$, $f_{A}[u]$ is a functional
defined on the gauge orbit of $A_{\mu }$. Let $\mathcal{A}$ be the space of
connections $A_{\mu }^{a}$ with finite Hilbert norm $||A||$, \textit{i.e.}
\begin{equation}
||A||^{2}=\mathrm{Tr}\int d^{4}x\,A_{\mu }A{_{\mu }=}\frac{1}{2}\int
d^{4}xA_{\mu }^{a}A_{\mu }^{a}<+\infty \;, \label{norm0}
\end{equation}
and let $\mathcal{U}$ be the space of local gauge transformations $u$ such
that the Hilbert norm $||u^{\dagger }\partial {u}||$ is finite too, namely
\begin{equation}
||u^{\dagger }\partial {u}||^{2}=\mathrm{Tr}\int d^{4}x\,\left(
u^{\dagger }\partial _{\mu }u\right) \left( u^{\dagger }\partial
_{\mu }u\right) <+\infty \;. \label{norm1}
\end{equation}
\noindent The following proposition holds
\cite{Zwanziger:1990tn,Dell'Antonio:1989jn,Dell'Antonio:1991xt,vanBaal:1991zw}
\begin{itemize}
\item  \underline{Proposition}\newline
The functional $f_{A}[u]$ achieves its absolute minimum on the gauge orbit
of $A_{\mu }$.
\end{itemize}
\noindent This proposition means that there exists a $h\in
\mathcal{U}$ such that
\begin{eqnarray}
\delta f_{A}[h] &=&0\;,  \label{impl0} \\
\delta ^{2}f_{A}[h] &\ge &0\;,  \label{impl1} \\
f_{A}[h] &\le &f_{A}[u]\;,\;\;\;\;\;\;\;\forall \,u\in
\mathcal{U}\;. \label{impl2}
\end{eqnarray}
The operator $A_{\min }^{2}$ is thus given by
\begin{equation}
A_{\min }^{2}=\min_{\left\{ u\right\} }\mathrm{Tr}\int
d^{4}x\,A_{\mu }^{u}A_{\mu }^{u}=f_{A}[h]\;.  \label{a2min}
\end{equation}
Let us give a look at the two conditions (\ref{impl0}) and
(\ref{impl1}). To evaluate $\delta f_{A}[h]$ and $\delta
^{2}f_{A}[h]$ we set\footnote{The case of the gauge group $SU(N)$
is considered here.}
\begin{equation}
v=he^{ig\omega }=he^{ig\omega ^{a}T^{a}}\;,  \label{set0}
\end{equation}
\begin{equation}
\left[ T^{a},T^{b}\right] =if^{abc}\;T^c\;,\;\;\;\;\;\mathrm{Tr}\left( T^{a}T^{b}\right) =%
\frac{1}{2}\delta ^{ab}\;,  \label{st000}
\end{equation}
where $\omega $ is an infinitesimal hermitian matrix and we
compute the linear and quadratic terms of the expansion of the
functional $f_{A}[v]$ in power series of $\omega $. Let us first
obtain an expression for $A_{\mu }^{v}$
\begin{eqnarray}
A_{\mu }^{v} &=&v^{\dagger }A_{\mu }v+\frac{i}{g}v^{\dagger }\partial _{\mu
}v  \nonumber \\
&=&e^{-ig\omega }h^{\dagger }A_{\mu }he^{ig\omega }+\frac{i}{g}e^{-ig\omega
}\left( h^{\dagger }\partial _{\mu }h\right) e^{ig\omega }+\frac{i}{g}%
e^{-ig\omega }\partial _{\mu }e^{ig\omega }  \nonumber \\
&=&e^{-ig\omega }A_{\mu }^{h}e^{ig\omega }+\frac{i}{g}e^{-ig\omega }\partial
_{\mu }e^{ig\omega }\;.  \label{orbit0}
\end{eqnarray}
Expanding up to the order $\omega ^{2}$, we get
\begin{eqnarray}
A_{\mu }^{v} &=&\left( 1-ig\omega -g^{2}\frac{\omega ^{2}}{2}\right) A_{\mu
}^{h}\left( 1+ig\omega -g^{2}\frac{\omega ^{2}}{2}\right) +\frac{i}{g}\left(
1-ig\omega -g^{2}\frac{\omega ^{2}}{2}\right) \partial _{\mu }\left(
1+ig\omega -g^{2}\frac{\omega ^{2}}{2}\right)  \nonumber \\
&=&\left( 1-ig\omega -g^{2}\frac{\omega ^{2}}{2}\right) \left( A_{\mu
}^{h}+igA_{\mu }^{h}\omega -g^{2}A_{\mu }^{h}\frac{\omega ^{2}}{2}\right) +
\nonumber \\
&+&\frac{i}{g}\left( 1-ig\omega -g^{2}\frac{\omega ^{2}}{2}\right) \left(
ig\partial _{\mu }\omega -\frac{g^{2}}{2}\left( \partial _{\mu }\omega
\right) \omega -\frac{g^{2}}{2}\omega \left( \partial _{\mu }\omega \right)
\right)  \nonumber \\
&=&A_{\mu }^{h}+igA_{\mu }^{h}\omega -\frac{g^{2}}{2}A_{\mu }^{h}\omega
^{2}-ig\omega A_{\mu }^{h}+g^{2}\omega A_{\mu }^{h}\omega -\frac{g^{2}}{2}%
\omega ^{2}A_{\mu }^{h}  \nonumber \\
&+&\frac{i}{g}\left( ig\partial _{\mu }\omega -\frac{g^{2}}{2}\left(
\partial _{\mu }\omega \right) \omega -\frac{g^{2}}{2}\omega \partial _{\mu
}\omega +g^{2}\omega \partial _{\mu }\omega \right) +O(\omega ^{3})\;,
\label{ex1}
\end{eqnarray}
from which it follows
\begin{equation}
A_{\mu }^{v}=A_{\mu }^{h}+ig[A_{\mu }^{h},\omega ]+\frac{g^{2}}{2}[[\omega
,A_{\mu }^{h}],\omega ]-\partial _{\mu }\omega +i\frac{g}{2}[\omega
,\partial _{\mu }\omega ]+O(\omega ^{3})\;,  \label{A0}
\end{equation}
We now evaluate
\begin{eqnarray}
f_{A}[v] &=&\mathrm{Tr}\int d^{4}xA_{\mu }^{u}A_{\mu }^{u}  \nonumber \\
&=&\mathrm{Tr}\int d^{4}x\,\left[ \left( A_{\mu }^{h}+ig[A_{\mu }^{h},\omega ]+\frac{%
g^{2}}{2}[[\omega ,A_{\mu }^{h}],\omega ]-\partial _{\mu }\omega +i\frac{g}{2%
}[\omega ,\partial _{\mu }\omega ]+O(\omega ^{3})\right) \times
\right.
\nonumber \\
& &\left. \left( A_{\mu }^{h}+ig[A_{\mu }^{h},\omega ]+\frac{g^{2}}{2}[%
[\omega ,A_{\mu }^{h}],\omega ]-\partial _{\mu }\omega +i\frac{g}{2}[\omega
,\partial _{\mu }\omega ]+O(\omega ^{3})\right) \right]  \nonumber \\
&=&\mathrm{Tr}\int d^{4}x\,\left\{ A_{\mu }^{h}A_{\mu
}^{h}+igA_{\mu }^{h}[A_{\mu
}^{h},\omega ]+g^{2}A_{\mu }^{h}\omega {A}_{\mu }^{h}\omega -\frac{g^{2}}{2}%
A_{\mu }^{h}A_{\mu }^{h}\omega ^{2}-\frac{g^{2}}{2}A_{\mu }^{h}\omega
^{2}A_{\mu }^{h}-A_{\mu }^{h}\partial _{\mu }\omega \right.  \nonumber \\
&+&\left. i\frac{g}{2}A_{\mu }^{h}[\omega ,\partial _{\mu }\omega
]+ig[A_{\mu }^{h},\omega ]A_{\mu }^{h}-g^{2}[A_{\mu }^{h},\omega
][A_{\mu }^{h},\omega ]-ig[A_{\mu }^{h},\omega ]\partial _{\mu
}\omega +g^{2}\omega
A_{\mu }^{h}\omega A_{\mu }^{h}\right.  \nonumber \\
&-&\left. \frac{g^{2}}{2}A_{\mu }^{h}\omega ^{2}A_{\mu }^{h}-\frac{g^{2}}{2}%
\omega ^{2}A_{\mu }^{h}A_{\mu }^{h}-\partial _{\mu }\omega A_{\mu
}^{h}-ig\partial _{\mu }\omega [A_{\mu }^{h},\omega ]+\partial _{\mu }\omega
\partial _{\mu }\omega +i\frac{g}{2}[\omega ,\partial _{\mu }\omega ]A_{\mu
}^{h}\right\} +O(\omega ^{3})\nonumber\\
 &=&f_{A}[h]-\mathrm{Tr}\int d^{4}x\,\left\{
A_{\mu }^{h},\partial _{\mu }\omega
\right\} +\mathrm{Tr}\int d^{4}x\,\left( g^{2}A_{\mu }^{h}\omega A_{\mu }^{h}\omega -%
\frac{g^{2}}{2}A_{\mu }^{h}A_{\mu }^{h}\omega ^{2}-\frac{g^{2}}{2}A_{\mu
}^{h}\omega ^{2}A_{\mu }^{h}\right.  \nonumber \\
&-&\left. g^{2}[A_{\mu }^{h},\omega ][A_{\mu }^{h},\omega
]+g^{2}\omega A_{\mu }^{h}\omega A_{\mu
}^{h}-\frac{g^{2}}{2}A_{\mu }^{h}\omega ^{2}A_{\mu
}^{h}-\frac{g^{2}}{2}\omega ^{2}A_{\mu }^{h}A_{\mu }^{h}\right)
+\mathrm{Tr}\int d^{4}x\,\left( \partial _{\mu }\omega \partial
_{\mu }\omega \right.
\nonumber \\
&+&\left. i\frac{g}{2}[\omega ,\partial _{\mu }\omega ]A_{\mu
}^{h}-ig\partial _{\mu }\omega [A_{\mu }^{h},\omega ]-ig[A_{\mu
}^{h},\omega ]\partial _{\mu }\omega +i\frac{g}{2}A_{\mu
}^{h}[\omega ,\partial _{\mu }\omega ]\right) +O(\omega ^{3})
\nonumber
\end{eqnarray}
\begin{eqnarray}
&=&f_{A}[h]+2\int {d^{4}x}\,tr\left( \omega \partial _{\mu
}{A}_{\mu }^{h}\right) +\int {d^{4}x}\,tr\left\{ 2g^{2}\omega
{A}_{\mu }^{h}\omega
A_{\mu }^{h}-2g^{2}A_{\mu }^{h}A_{\mu }^{h}\omega ^{2}\right.  \nonumber \\
&-&\left. g^{2}\left( A_{\mu }^{h}\omega -\omega {A}_{\mu }^{h}\right)
\left( A_{\mu }^{h}\omega -\omega {A}_{\mu }^{h}\right) \right\} +\int {%
d^{4}x}\,tr\left( \partial _{\mu }\omega \partial _{\mu }\omega +i\frac{g}{2}%
\omega \partial _{\mu }\omega {A}_{\mu }^{h}-i\frac{g}{2}\partial _{\mu
}\omega \omega {A}_{\mu }^{h}\right.  \nonumber \\
&-&\left. ig\partial _{\mu }\omega {A}_{\mu }^{h}\omega +ig\partial _{\mu
}\omega \omega {A}_{\mu }^{h}-igA_{\mu }^{h}\omega \partial _{\mu }\omega
+ig\omega {A}_{\mu }^{h}\partial _{\mu }\omega +i\frac{g}{2}A_{\mu
}^{h}\omega \partial _{\mu }\omega -i\frac{g}{2}A_{\mu }^{h}\partial _{\mu
}\omega \omega \right) +O(\omega ^{3})  \nonumber \\
&=&f_{A}[h]+2\mathrm{Tr}\int d^{4}x\left( \,\omega \partial _{\mu
}A_{\mu }^{h}\right) +\mathrm{Tr}\int d^{4}x\,\left( \partial
_{\mu }\omega
\partial _{\mu }\omega +ig\omega \partial _{\mu }\omega {A}_{\mu
}^{h}-ig\partial _{\mu
}\omega \omega {A}_{\mu }^{h}\right.  \nonumber \\
&-&\left. 2ig\partial _{\mu }\omega A_{\mu }^{h}\omega +2ig\partial
_{\mu }\omega \omega A_{\mu }^{h}\right) +O(\omega ^{3})\;.
\end{eqnarray}
Thus
\begin{eqnarray}
f_{A}[v] &=&f_{A}[h]+2\mathrm{Tr}\int d^{4}x\,\left( \omega
\partial _{\mu }A_{\mu }^{h}\right) +\mathrm{Tr}\int d^{4}x\,\left(
\partial _{\mu }\omega \partial _{\mu }\omega +ig\omega \partial
_{\mu }\omega A_{\mu }^{h}-ig\partial _{\mu
}\omega \omega A_{\mu }^{h}\right.  \nonumber \\
&-&\left. ig\left( \partial _{\mu }\omega \right) A_{\mu }^{h}\omega
+ig\left( \partial _{\mu }\omega \right) \omega A_{\mu }^{h}\right)
+O(\omega ^{3})  \nonumber \\
&=&f_{A}[h]+2\mathrm{Tr}\int d^{4}x\,\left( \omega \partial _{\mu
}A_{\mu }^{h}\right) +\mathrm{Tr}\int d^{4}x\,\left\{ \partial
_{\mu }\omega \left(
\partial _{\mu }\omega -ig\left[ A_{\mu }^{h},\omega \right]
\right) \right\} +O(\omega ^{3})\;. \nonumber  \\ \label{f1}
\end{eqnarray}
Finally
\begin{equation}
f_{A}[v]=f_{A}[h]+2\mathrm{Tr}\int d^{4}x\,\left( \omega \partial
_{\mu }A_{\mu }^{h}\right) -\mathrm{Tr}\int d^{4}x\,\omega
\partial _{\mu }D_{\mu }(A^{h})\omega +O(\omega ^{3})\;,
\label{func2}
\end{equation}
so that
\begin{eqnarray}
\delta f_{A}[h] &=&0\;\;\;\Rightarrow \;\;\;\partial _{\mu }A_{\mu
}^{h}\;=\;0\;,  \nonumber \\
\delta ^{2}f_{A}[h] &>&0\;\;\;\Rightarrow \;\;\;-\partial _{\mu }D{_{\mu }(}%
A^{h}{)}\;>\;0\;.  \label{func3}
\end{eqnarray}
We see therefore that the set of field configurations fulfilling conditions (%
\ref{func3}), \textit{i.e.} defining relative minima of the functional $%
f_{A}[u]$, belong to the so called Gribov region $\Omega $, which is defined
as
\begin{equation}
\Omega =\left.\{A_{\mu }\right|\partial _{\mu }A_{\mu
}=0\;\mathrm{and}\;-\partial _{\mu }D_{\mu }(A)>0\}\;.
\label{gribov0}
\end{equation}
Let us proceed now by showing that the transversality condition,
$\partial
_{\mu }A_{\mu }^{h}=0$, can be solved for $h=h(A)$ as a power series in $%
A_{\mu }$. We start from
\begin{equation}
A_{\mu }^{h}=h^{\dagger }A_{\mu }h+\frac{i}{g}h^{\dagger }\partial _{\mu
}h\;,  \label{Ah0}
\end{equation}
with
\begin{equation}
h=e^{ig\phi }=e^{ig\phi ^{a}T^{a}}\;.  \label{h0}
\end{equation}
Let us expand $h$ in powers of $\phi $
\begin{equation}
h=1+ig\phi -\frac{g^{2}}{2}\phi ^{2}+O(\phi ^{3})\;.  \label{hh1}
\end{equation}
From equation (\ref{Ah0}) we have
\begin{equation}
A_{\mu }^{h}=A_{\mu }+ig[A_{\mu },\phi ]+g^{2}\phi A_{\mu }\phi -\frac{g^{2}%
}{2}A_{\mu }\phi ^{2}-\frac{g^{2}}{2}\phi ^{2}A_{\mu }-\partial _{\mu }\phi
+i\frac{g}{2}[\phi ,\partial _{\mu }]+O(\phi ^{3})\;.  \label{A1}
\end{equation}
Thus, condition $\partial _{\mu }A_{\mu }^{h}=0$, gives
\begin{eqnarray}
\partial ^{2}\phi &=&\partial _{\mu }A+ig[\partial _{\mu }A_{\mu },\phi
]+ig[A_{\mu },\partial _{\mu }\phi ]+g^{2}\partial _{\mu }\phi A_{\mu }\phi
+g^{2}\phi \partial _{\mu }A_{\mu }\phi +g^{2}\phi A_{\mu }\partial _{\mu
}\phi   \nonumber \\
&-&\frac{g^{2}}{2}\partial _{\mu }A_{\mu }\phi ^{2}-\frac{g^{2}}{2}A_{\mu
}\partial _{\mu }\phi \phi -\frac{g^{2}}{2}A_{\mu }\phi \partial _{\mu }\phi
-\frac{g^{2}}{2}\partial _{\mu }\phi \phi A_{\mu }-\frac{g^{2}}{2}\phi
\partial _{\mu }\phi A_{\mu }-\frac{g^{2}}{2}\phi ^{2}\partial _{\mu }A_{\mu
}  \nonumber \\
&+&i\frac{g}{2}[\phi ,\partial ^{2}\phi ]+O(\phi ^{3})\;.  \label{hh2}
\end{eqnarray}
This equation can be solved iteratively for $\phi $ as a power series in $%
A_{\mu }$, namely
\begin{equation}
\phi =\frac{1}{\partial ^{2}}\partial _{\mu }A_{\mu }+i\frac{g}{\partial ^{2}%
}\left[ \partial A,\frac{\partial A}{\partial ^{2}}\right] +i\frac{g}{%
\partial ^{2}}\left[ A_{\mu },\partial _{\mu }\frac{\partial A}{\partial ^{2}%
}\right] +\frac{i}{2}\frac{g}{\partial ^{2}}\left[ \frac{\partial A}{%
\partial ^{2}},\partial A\right] +O(A^{3})\;,  \label{phi0}
\end{equation}
so that
\begin{eqnarray}
A_{\mu }^{h} &=&A_{\mu }-\frac{1}{\partial ^{2}}\partial _{\mu }\partial A-ig%
\frac{\partial _{\mu }}{\partial ^{2}}\left[ A_{\nu },\partial _{\nu }\frac{%
\partial A}{\partial ^{2}}\right] -i\frac{g}{2}\frac{\partial _{\mu }}{%
\partial ^{2}}\left[ \partial A,\frac{1}{\partial ^{2}}\partial A\right]
\nonumber \\
&+&ig\left[ A_{\mu },\frac{1}{\partial ^{2}}\partial A\right] +i\frac{g}{2}%
\left[ \frac{1}{\partial ^{2}}\partial A,\frac{\partial _{\mu }}{\partial
^{2}}\partial A\right] +O(A^{3})\;.  \label{minn2}
\end{eqnarray}
Expression (\ref{minn2}) can be written in a more useful way,
given in eq.(\ref{min0}). In fact
\begin{eqnarray}
A_{\mu }^{h} &=&\left( \delta _{\mu \nu }-\frac{\partial _{\mu }\partial
_{\nu }}{\partial ^{2}}\right) \left( A_{\nu }-ig\left[ \frac{1}{\partial
^{2}}\partial A,A_{\nu }\right] +\frac{ig}{2}\left[ \frac{1}{\partial ^{2}}%
\partial A,\partial _{\nu }\frac{1}{\partial ^{2}}\partial A\right] \right)
+O(A^{3})  \nonumber \\
&=&A_{\mu }-ig\left[ \frac{1}{\partial ^{2}}\partial A,A_{\mu }\right] +%
\frac{ig}{2}\left[ \frac{1}{\partial ^{2}}\partial A,\partial _{\mu }\frac{1%
}{\partial ^{2}}\partial A\right] -\frac{\partial _{\mu }}{\partial ^{2}}%
\partial A+ig\frac{\partial _{\mu }}{\partial ^{2}}\partial _{\nu }\left[
\frac{1}{\partial ^{2}}\partial A,A_{\nu }\right]   \nonumber \\
&-&i\frac{g}{2}\frac{\partial _{\mu }}{\partial ^{2}}\partial _{\nu }\left[
\frac{\partial A}{\partial ^{2}},\frac{\partial _{\nu }}{\partial ^{2}}%
\partial A\right] +O(A^{3})  \nonumber \\
&=&A_{\mu }-\frac{\partial _{\mu }}{\partial ^{2}}\partial A+ig\left[ A_{\mu
},\frac{1}{\partial ^{2}}\partial A\right] +\frac{ig}{2}\left[ \frac{1}{%
\partial ^{2}}\partial A,\partial _{\mu }\frac{1}{\partial ^{2}}\partial
A\right] +ig\frac{\partial _{\mu }}{\partial ^{2}}\left[ \frac{\partial
_{\nu }}{\partial ^{2}}\partial A,A_{\nu }\right]   \nonumber \\
&+&i\frac{g}{2}\frac{\partial _{\mu }}{\partial ^{2}}\left[ \frac{\partial A%
}{\partial ^{2}},\partial A\right] +O(A^{3})  \label{hhh3}
\end{eqnarray}
which is precisely expression (\ref{minn2}). The transverse field
given in eq.(\ref {min0}) enjoys the property of being gauge
invariant order by order in the
coupling constant $g$. Let us work out the transformation properties of $%
\phi _{\nu }$ under a gauge transformation
\begin{equation}
\delta A_{\mu }=-\partial _{\mu }\omega +ig[A_{\mu },\omega ]\;.
\label{gauge3}
\end{equation}
We have, up to the order $O(g^{2})$,
\begin{eqnarray}
\delta \phi _{\nu } &=&-\partial _{\nu }\omega +ig\left[ \frac{1}{\partial
^{2}}\partial A,\partial _{\nu }\omega \right] -i\frac{g}{2}\left[ \omega
,\partial _{\nu }\frac{1}{\partial ^{2}}\partial A\right] -i\frac{g}{2}%
\left[ \frac{\partial A}{\partial ^{2}},\partial _{\nu }\omega \right]
+O(g^{2})  \nonumber \\
&=&-\partial _{\nu }\omega +i\frac{g}{2}\left[ \frac{1}{\partial ^{2}}%
\partial A,\partial _{\nu }\omega \right] +i\frac{g}{2}\left[ \partial _{\nu
}\frac{1}{\partial ^{2}}\partial A,\omega \right] +O(g^{2})\;.  \label{gg2}
\end{eqnarray}
Therefore
\begin{equation}
\delta \phi _{\nu }=-\partial _{\nu }\left( \omega -i\frac{g}{2}\left[ \frac{%
\partial A}{\partial ^{2}},\omega \right] \right) +O(g^{2})\;,  \label{phi1}
\end{equation}
from which the gauge invariance of $A_{\mu }^{h}$ is established.\newline
\newline
Finally, let us work out the expression of $A_{\mathrm{min}}^{2}$ as a power
series in $A_{\mu }$.
\begin{eqnarray}
A_{\mathrm{min}}^{2} &=&\mathrm{Tr}\int d^{4}x\,A_{\mu }^{h}A_{\mu }^{h}  \nonumber \\
&=&\mathrm{Tr}\int d^{4}x\,\left[ \phi _{\mu }\left( \delta _{\mu \nu }-\frac{%
\partial _{\mu }\partial _{\nu }}{\partial ^{2}}\right) \phi _{\nu }\right]
\nonumber \\
&=&\mathrm{Tr}\int d^{4}x\,\left[ \left( A_{\mu }-ig\left[ \frac{1}{\partial ^{2}}%
\partial A,A_{\mu }\right] +\frac{ig}{2}\left[ \frac{1}{\partial ^{2}}%
\partial A,\partial _{\mu }\frac{1}{\partial ^{2}}\partial A\right] \right)
\times \right.  \nonumber \\
&&\left.  \left( \delta _{\mu \nu }-\frac{\partial _{\mu }\partial
_{\nu }}{\partial ^{2}}\right) \left( A_{\nu }-ig\left[
\frac{1}{\partial
^{2}}\partial A,A_{\nu }\right] +\frac{ig}{2}\left[ \frac{1}{\partial ^{2}}%
\partial A,\partial _{\nu }\frac{1}{\partial ^{2}}\partial A\right] \right)
\right]  \nonumber \\
&=&\frac{1}{2}\int d^{4}x\left[ A_{\mu }^{a}\left( \delta _{\mu \nu }-\frac{%
\partial _{\mu }\partial _{\nu }}{\partial ^{2}}\right) A_{\nu
}^{a}-2gf^{abc}\frac{\partial _{\nu }\partial A^{a}}{\partial ^{2}}\frac{%
\partial A^{b}}{\partial ^{2}}A_{\nu }^{c}-gf^{abc}A_{\nu }^{a}\frac{%
\partial A^{b}}{\partial ^{2}}\frac{\partial _{\nu }\partial A^{c}}{\partial
^{2}}\right] +O(A^{4})\;.  \nonumber \\
&&  \label{a2em}
\end{eqnarray}
leading to the result quoted in eq.(\ref{min1}).\newline\newline
We conclude this Appendix by noting that, due to gauge invariance, $A_{%
\mathrm{min}}^{2}$ can be rewritten in a manifestly invariant way
in terms of $F_{\mu \nu }$ and the covariant derivative $D_{\mu }$
\cite{Zwanziger:1990tn}, see eq.(\ref{zzw}).


\section{Propagators of the elementary fields}\label{appb}

In order to evaluate the tree-level two-point functions of the theory, we start from the local action 
\begin{equation}
S = S_{FP}+
\int d^4x \left(\tau^{a}\,\partial_{\mu}A^{h,a}_{\mu}
+\frac{m^{2}}{2}\,A^{h,a}_{\mu}A^{h,a}_{\mu}\right) + S_{IRR}    \label{pa}
\end{equation} 
where $S_{FP}$ is the Faddeev-Popov term of the linear covariant gauges, eq.\eqref{S_FP}, and $S_{IRR}$ stands for the BRST-invariant infrared regularizing mass term for 
the Stueckelberg field, namely 
\begin{eqnarray}
S_{IRR}&=&\int d^4x \frac{1}{2} s\left(\rho\xi^a\xi^a\right)
=\int d^4x \left( \frac{1}{2}M^4\xi^a\xi^a+\rho\xi^ac^a  \right)
\end{eqnarray}
From the quadratic part of expression \eqref{pa}, one finds the following set of tree-level propagators in momentum space
\begin{eqnarray}
\langle A^a_{\mu}(p)A^b_{\nu}(-p)\rangle &=& \frac{1}{p^2+m^2} \delta^{ab}\mathcal{P}_{\mu\nu}
+\frac{\alpha}{p^2}\frac{p_{\mu}p_{\nu}}{p^2} 
\\
\langle A_{\mu}^a(p)b^b(-p)\rangle&=& -\frac{p^2}{p^4+\alpha M^4}\delta^{ab}p_{\mu}
\\
\langle A_{\mu}^a(p)\xi^b(-p)\rangle&=& i\frac{\alpha\delta^{ab}}{p^4+\alpha M^4}p_{\mu}\\
\langle A_{\mu}^a(p)\tau^b(-p)\rangle&=&-i\frac{\alpha M^4}{p^2(p^4+\alpha M^4)}p_{\mu}\delta^{ab}
\end{eqnarray}
\begin{eqnarray}
\langle b^a(p)b^b(-p)\rangle &=&\frac{M^4}{p^4+\alpha M^4}\delta^{ab}
\\
\langle b^a(p)\xi^b(-p)\rangle&=&i\frac{p^2\delta^{ab}}{p^4+\alpha M^4}
\\
\langle b^a(p)\tau^b(-p)\rangle&=&-i\frac{M^4}{p^2}\delta^{ab}
\\
\langle\bar{c}^a(p)A^b_{\mu}(-p)\rangle &=&-i\frac{\rho\,\alpha}{p^2(p^4+\alpha M^4)}\delta^{ab}p_{\mu}\\
\langle\bar{c}^a(p)b^b(-p)\rangle &=&i\frac{\rho}{p^4+\alpha M^4}\delta^{ab}\\
\langle\bar{c}^a(p)\tau^b(-p)\rangle &=&\frac{\rho}{p^4+\alpha M^4}\delta^{ab}\\
\langle\bar{c}^a(p)\xi^b(-p)\rangle &=&\frac{\rho\,\alpha}{p^2(p^4+\alpha M^4)}\delta^{ab}\\
\langle\xi^a(p)\xi^b(-p)\rangle&=&\frac{\alpha\delta^{ab}}{p^4+\alpha M^4}
\\
\langle\xi^a(p)\tau^b(-p)\rangle &=&\frac{p^2}{p^4+\alpha M^4}\delta^{ab}
\\
\langle\tau^a(p)\tau^b(-p)\rangle&=&-\left(\frac{m^2(p^4-\alpha M^4)+M^4p^2}{p^2(p^4+\alpha M^4)}
\right)\delta^{ab}   \\
\langle\bar{c}^a(p)c^b(-p)\rangle &=&\frac{1}{p^2}\delta^{ab} \;,
\end{eqnarray}
with $\mathcal{P}_{\mu\nu} = \left(\delta_{\mu\nu} - \frac{p_\mu p_\nu}{p^2} \right)$ being the transverse projector.
 All other propagators which have not been listed above are vanishing. Let us also remind that the parameters
$M$ and $\rho$ which regularize the propagation of the Stueckelberg field in the infrared have to be set to zero at
the end of any actual calculation.

\end{document}